\documentclass[pre,aps,superscriptaddress,showpacs,floatfix,notitlepage]{revtex4-1}
\usepackage{amssymb,mathrsfs,amsfonts}
\usepackage{amsmath}
\usepackage{graphicx}
\usepackage{epstopdf}
\usepackage{nicefrac}
\usepackage{listings}
\usepackage{amsmath}
\usepackage{upgreek}
\usepackage{multirow}
\usepackage{color}

\newcommand{\A}{{\cal A}}
\newcommand{\B}{{\cal B}}

\newcommand{\RN}[1]{%
  \textup{\uppercase\expandafter{\romannumeral#1}}%
}

\begin{document}
\title{The Discontinuous Asymptotic Telegrapher's Equation ($P_1$) Approximation} 
\author{Avner P. Cohen}
\email{avnerco@gmail.com}
\affiliation{Department of Physics, Nuclear Research Center-Negev, P.O. Box 9001, Beer-Sheva 84190, ISRAEL}
\author{Roy Perry}
\affiliation{Department of Physics, Ben-Gurion University, Beer-Sheva 84105, ISRAEL}
\author{Shay I. Heizler}
\email{highzlers@walla.co.il}
\affiliation{Department of Physics, Nuclear Research Center-Negev, P.O. Box 9001, Beer-Sheva 84190, ISRAEL}


\begin{abstract}
Modeling the propagation of radiative heat-waves in optically thick material using a diffusive approximation is a well-known problem. In 
optically thin material, classic methods, such as classic diffusion or classic $P_1$, yield the wrong heat wave propagation behavior, and higher order approximation might be required, making the solution harder to obtain. The asymptotic $P_1$ approximation [Heizler, {\em NSE} 166, 17 (2010)] yields the correct particle velocity but fails to model the correct behavior in highly anisotropic media, such as problems that involve sharp boundary between media or strong sources. However, the solution for the two-region Milne problem of two adjacent half-spaces divided by a sharp boundary, yields a discontinuity in the asymptotic solutions, that makes it possible to solve steady-state problems, especially in neutronics. In this work we expand the time-dependent asymptotic $P_1$ approximation to a highly anisotropic media, using the discontinuity jump conditions of the energy density, yielding a modified discontinuous $P_1$ equations in general geometry. We introduce numerical solutions for two fundamental benchmarks in plane symmetry. 
The results thus obtained are more accurate than those attained by other methods,
such as Flux-Limiters or Variable Eddington Factor. 
\end{abstract}

\pacs{} \maketitle

\section{Introduction}
\label{s1}

Radiation heat waves (Marshak waves) play important roles in many high energy density physical phenomena, for example in inertial confinement fusion (ICF) and in astrophysical and laboratory
plasmas~\cite{Zeldovich2002,Lindl2004,Rosen1996,Mihalis1984}. This problem has long been a subject of theoretical astrophysics research 
~\cite{Chandrasekhar1935,Milne1921}, and of experimental studies testing radiative-hydrodynamics macroscopic modeling~\cite{Back2000,Moore2015}.
Specifically, the propagating radiative Marshak waves in optically thick media are well described by a simple local thermodynamic equilibrium (LTE) diffusion model,
yielding self-similar solutions of both supersonic and subsonic regimes~\cite{Marshak1958,Pakula1985,Shussman2015,Malka2016}. However, in optically
thin media, the diffusion limit fails to describe the exact physical behavior of the problem. In the general case, the propagation of the radiation is modeled
via the Boltzmann (transport) equation for photons, coupled to the matter via the energy balance equation.
In the gray (mono-energetic) radiation case the equation is:
\begin{equation}
\begin{split}
\frac{1}{c}\frac{\partial I(\hat{\Omega},\vec{r},t)}{\partial t}+\hat{\Omega}\cdot \vec{\nabla}I(\hat{\Omega},\vec{r},t)+
& \left(\sigma_{a}(T_m(\vec{r},t))+\sigma_{s}(T_m(\vec{r},t))\right)I(\hat{\Omega},\vec{r},t)=\\
& \sigma_{a}(T_m(\vec{r},t)){B}(T_m(\vec{r},t))+\\
& \frac{\sigma_{s}(T_m(\vec{r},t))}{4\pi}\int_{4\pi}I(\hat{\Omega},\vec{r},t)d\hat{\Omega}+
S(\hat{\Omega},\vec{r},t)
\end{split}
\label{Boltz}
\end{equation}
where $I(\hat{\Omega},\vec{r},t)$ is the specific intensity of radiation at position $\vec{r}$ propagating in the $\hat{\Omega}$  direction at time $t$. $B(T_m(\vec{r},t))$ is the thermal material energy, where $T_m(\vec{r},t)$ is the material temperature, $c$ is the speed of light and $S(\hat{\Omega},\vec{r},t)$ is
an external radiation source.
$\sigma_{a}(T_m(\vec{r},t))$ and $\sigma_{s}(T_m(\vec{r},t))$ are the absorption (opacity) and scattering cross-sections respectively.
In this paper we focus on the gray case, when the expansion to multi-energy approximation is straightforward~\cite{Pomraning1973}. 
Along with the equation for the radiation energy, the complementary equation for the material is:
\begin{equation}
\frac{C_v(T_m(\vec{r},t))}{c}\frac{\partial T_m(\vec{r},t)}{\partial t}=
\sigma_{a}(T_m(\vec{r},t))\left(\frac{1}{c}\int_{4\pi}{I(\hat{\Omega},\vec{r},t)d\hat{\Omega}}-aT_m^4(\vec{r},t)\right)
\label{Matter1}
\end{equation}
where $C_v(T_m(\vec{r},t))$ is the heat capacity of the material.

Solving the transport equation is complicated, especially in multi-dimensions, where an exact solution is hard to obtain. The $P_N$ approximation, which decomposes $I(\hat{\Omega},\vec{r},t)$ to its first $N$ angular moments (defines $N$ coupled equations, assuming the $P_N$ closure), and
the $S_N$ method (the transport equation in $N$ discrete ordinates), are deterministic methods, and they are both exact when $N\to\infty$~\cite{Pomraning1973}.
Alternatively, a statistical implicit Monte Carlo (IMC) approach can also be used~\cite{IMC}, which is exact when the number of particles (histories) goes to infinity.
Although these three methods approach the exact solution, their application requires extensive numerical calculations that might be difficult to carry out, especially in multi-dimensions.
Hence, there is an extensive body of literature dealing with the search for approximate models which will be relatively easy to simulate, and yet produce solutions that are close to the exact problem (for example, see~\cite{Olson1999,Su2001}).

The classical (Eddington) diffusion theory, as a specific case of the $P_1$ is relatively easy to solve and is commonly used~\cite{Pomraning1973,Zeldovich2002,Mihalis1984}.
The diffusion equation is parabolic, and thus yields infinite particle velocities. The full $P_1$ equations, that give rise to the Telegrapher's equation, has a
hyperbolic form, but with an incorrect finite velocity, $c/\sqrt{3}$~\cite{Heizler2010}. Possible solutions, such as flux-limiters (FL) solution (in the form of a non-linear diffusion notation), or Variable Eddington Factor (VEF) approximations (in the form of full $P_1$ equations), yielding a gradient-dependent nonlinear diffusion coefficients (or a gradient-dependent Eddington factor), are harder to solve, especially in multi-dimensions~\cite{Winslow1968,Minerbo1978,Pomraning_survey,LevermorePomraning1981,Pomraning1982,LEVERMORE1983,Pomraning1984,Olson1999,Su2001}.

In previous work, Heizler~\cite{Heizler2010} offered a modified $P_1$ approximation, based on the asymptotic derivation (both in space and time), the asymptotic $P_1$ approximation (or the asymptotic Telegrapher's equation approximation)~\cite{Heizler2010}. In steady state, it tends to the well-known asymptotic diffusion approximation~\cite{Frankel1953,Case1953,Pomraning1973}. This approximation shares similar asymptotic behavior with the $SP_2$ approximation in highly isotropic problems~\cite{Ravetto_Heizler2012}. It was tested in radiation problems under the {\em LTE} assumption, yielding relatively good results, especially near the tails, but also producing significant deviations in the regions where the material and radiation temperatures differ significantly~\cite{Heizler2012}.

However, when the radiation intensity is highly anisotropic, for example near a sharp boundary between two different media or near strong sources, the asymptotic $P_1$ 
results, are almost as poor as the classic $P_1$ or asymptotic diffusion approximations. Similar problem occurred in neutronics, with a sharp boundary of two different media, such as reactor-reflector problems~\cite{doyas_koponen2}. This problem can be corrected by using the exact solution to obtain the exact scalar flux and the neutron current, on the boundary between the two media, yielding a discontinuous asymptotic diffusion theory~\cite{Korn1967,mccormick1,mccormick2,mccormick3,ganapol_pomraning}. This correction is the two-region extension to the classic radiative transfer Milne problem~\cite{Milne1921,Zeldovich2002}, that has its origins in the attempt to calculate the distribution of light emitted from the photosphere of a star. The problem can be solved where the star is modeled as a semi-infinite half-space (with a vacuum boundary condition). By this correction, the problem of steady-state critical values in reactor-reflector problems is accurately modeled~\cite{doyas_koponen2}. We note that Zimmerman~\cite{zimmerman1979} offered an approximate version of this solution, based on the two-region Marshak-like boundary condition~\cite{Pomraning1973}, in order to adjust the different zones. In his approach the scalar flux has a discontinuity on the boundary, but the neutron current is continuous (and thus conserves particles).
  
In this work we offer a time-dependent version of this approach, i.e.~expanding the asymptotic $P_1$ approximation to a non-homogeneous space problem. By assuming that the energy density (the zero's moment of the specific intensity $I(\hat{\Omega},\vec{r},t)$) has a discontinuity, we derive the {\em discontinuous asymptotic Telegrapher's equation ($P_1$) approximation}. Our new method will be compared to other known diffusion and flux-limiter approximations, as well as the $P_1$ approximation and the VEF approximations in two basic and important problems: The Su-Olson (constant opacity) benchmark~\cite{SuOlson1996,SuOlson1999}, and the nonlinear-opacity Olson's benchmark~\cite{Olson1999}. It is important to note that the extension of the discontinuous asymptotic $P_1$ approximation is also straightforward for neutronics.

The present paper is structured in the following manner: first, in Sec.~\ref{s2} we will introduce common approximations for the Boltzmann equation. In Sec.~\ref{s6} we present the derivation of the discontinuous asymptotic Telegraphers equation ($P_1$) approximation. Next, in Sec.~\ref{results} the various approximations will be tested in the well-known radiation benchmarks. In Sec.~\ref{alpha_beta_sec} we examine another version of a discontinuous $P_1$ approximation, forcing a discontinuity in both energy density and radiation flux. A short discussion is presented in Sec. \ref{discussion}.

\section{Approximate models for the Radiative Transfer Equation}
\label{s2}

The first two angular moments of the specific intensity $I(\hat{\Omega},\vec{r},t)$ can be expressed as:
\begin{equation}
E(\vec{r},t)=\frac{1}{c}\int_{4\pi}{I(\hat{\Omega},\vec{r},t)d\hat{\Omega}}
\label{Edf}
\end{equation}
\begin{equation}
\vec{F}(\vec{r},t)=\int_{4\pi}{ I(\hat{\Omega},\vec{r},t)\hat{\Omega}d\hat{\Omega}}
\label{Fdf}
\end{equation}
where $E(\vec{r},t)$ is the energy density, and $\vec{F}(\vec{r},t)$ is the radiation flux.

Integration Eq.~\ref{Boltz} over all solid angle $\int{{d}\hat{\Omega}}$ yields {\em the conservation law}:
\begin{equation}
\\ \frac{1}{c}\frac{\partial E(\vec{r},t)}{\partial t}+
\\ \frac{1}{c} \nabla\cdot \vec{F}(\vec{r},t)=\sigma_{a}(T_m(\vec{r},t))\left(\int_{4\pi}{\frac{B(\vec{r},t)}{c}}d\hat{\Omega}-E(\vec{r},t)\right)+\frac{S(\vec{r},t)}{c}
\label{Rad1}
\end{equation}
Integration Eq.~\ref{Boltz} with $\int{\hat{\Omega{d}}\hat{\Omega}}$ yields:
\begin{equation}
\\ \frac{1}{c}\frac{\partial \vec{F}(\vec{r},t)}{\partial t}+
c\vec{\nabla}\cdot\int_{4\pi}{ I(\hat{\Omega},\vec{r},t)\hat{\Omega}\hat{\Omega}d\Omega}+\sigma_{t}(T_m(\vec{r},t)) F(\vec{r},t)=0
\label{Rad2}
\end{equation}
when $\sigma_{t}(T_m(\vec{r},t))=\sigma_a(T_m(\vec{r},t))+\sigma_s(T_m(\vec{r},t))$ is the total cross-section.
Eqs.~\ref{Rad1} and~\ref{Rad2} are exact equations. In these equations there are $3$ unknown moments of $I(\hat{\Omega},\vec{r},t)$, but only two
equations. Hench, we have to assume a closure for this moments representation, i.e.~to introduce an approximation for the third moment: 
$\int_{4\pi}{ I(\hat{\Omega},\vec{r},t)\hat{\Omega}\hat{\Omega}d\Omega}$. In the following we introduce a set of approximations that retain the conservation law (Eq.~\ref{Rad1}) (allowing energy conservation), while an approximation is introduced for Eq.~\ref{Rad2} (and for the third moment). 

\subsection{The Classic Diffusion and $P_1$ (Telegrapher's Equation) Approximations}
\label{s3}
The classic diffusion (or the classic Eddington) approximation (which is a simplification of the $P_1$ approximation) is the most well-known approximation
for the Boltzmann (transport) equation~\cite{Pomraning1973} and is extensively
used, especially in radiative transfer equation (RTE).

In the derivation of the $P_1$ approximation, one assumes that $I(\hat{\Omega},\vec{r},t)$
is a sum of its first two moments. Therefore the third moment can be approximated as $\int_{4\pi}{ I(\hat{\Omega},\vec{r},t)\hat{\Omega}\hat{\Omega}d\Omega}\approx E(\vec{r},t)/3$. In this case, Eq.~\ref{Rad2} takes this form:
\begin{equation} 
\frac{1}{c} \frac{\partial \vec{F}(\vec{r},t)}{\partial t}+\frac{c}{3}\vec{\nabla}E(\vec{r},t)+\sigma_{t}(T_m(\vec{r},t))\vec{F}(\vec{r},t)=0
\label{Rad2Class}
\end{equation}
Eqs.~\ref{Rad1} and~\ref{Rad2Class}, defining the $P_1$ approximation, are a set of two closed equations for $E(\vec{r},t)$ and $\vec{F}(\vec{r},t)$, coupled with the material energy equation, Eq.~\ref{Matter1}.  

If the derivative of the energy flux $\vec{F}(\vec{r},t)$ with respect to time inside Eq.~\ref{Rad2Class} is negligible, a form of a Fick's law is obtained:
\begin{equation}
\vec{F}(\vec{r},t)=-cD(\vec{r},t)\vec{\nabla}E(\vec{r},t),
\label{ficks}
\end{equation}
where $D(\vec{r},t)=1/\left[3\sigma_t(T_m(\vec{r},t))\right]$.
Substituting Eq.~\ref{ficks} in Eq.~\ref{Rad1} gives a diffusion equation:
\begin{equation}
\\ \frac{1}{c}\frac{\partial E(\vec{r},t)}{\partial t}-
\\ \vec{\nabla}\left({D(\vec{r},t)}\vec{\nabla}E(\vec{r},t)\right)=
\\ \sigma_{a}(T_m(\vec{r},t))\left(\frac{B(\vec{r},t)}{c}-E(\vec{r},t)\right)+\frac{S(\vec{r},t)}{c}
\label{diff}
\end{equation}
We note that the classic diffusion approximation
yields a wrong time-description due to its parabolic nature;
the diffusion approximation yields an infinite particle velocity. 
The full $P_1$ approximation (Eqs.~\ref{Rad2Class} and~\ref{Rad1}) can be re-formulated in a hyperbolic form:
\begin{align}
\label{telegraphers}
& \frac{1}{c\sigma_t(T_m(\vec{r},t))}\frac{\partial^2 E(\vec{r},t)}{\partial t^2}-
\frac{c}{3}\cdot\nabla\frac{1}{\sigma_t(T_m(\vec{r},t))}\nabla{E(\vec{r},t)}+
\frac{\partial E(\vec{r},t)}{\partial t}+ \nonumber \\
&\frac{1}{\sigma_t(T_m(\vec{r},t))}\frac{\partial \left(\sigma_a(T_m(\vec{r},t)) E(\vec{r},t)\right)}{\partial t} 
+\sigma_a(T_m(\vec{r},t))c E(\vec{r},t)= \\
& 4\pi\sigma_a(T_m(\vec{r},t))B(\vec{r},t)+S(\vec{r})+\frac{4\pi}{c\sigma_t(T_m(\vec{r},t))}\frac{\partial \left(\sigma_a(T_m(\vec{r},t)) B(\vec{r},t)\right)}{\partial t}+
\frac{1}{\sigma_t(T_m(\vec{r},t))}\frac{\partial S(\vec{r},t)}{\partial t}  \nonumber
\end{align}
The equation is developed under the assumption
that both time derivative of $\vec{F}(\vec{r},t)$ and the opacity spatial change are small enough, so the $\vec{\nabla}\left(\frac{1}{\sigma_t(T_m(\vec{r},t))}\right)\cdot\frac{\partial \vec{F}(\vec{r},t)}{\partial t}$ term, can be neglected ~\cite{Heizler2010,Heizler2012}.
This equation is called the Telegrapher's equation, and it combines both the second and the first derivative of the energy density with respect to time. 
The particle velocity in the classic $P_1$ approximation is too small, $c/\sqrt{3}$~\cite{Heizler2010,Heizler2012}, unlike the classic diffusion particle velocity which is 
too fast.

\subsection{Flux-limiter diffusion and Variable Eddington factor approximations}
\label{s4}
The parabolic nature of the diffusion approximation can be corrected by using
a nonlinear diffusion coefficient; flux-limited diffusion coefficient~\cite{Pomraning_survey,LEVERMORE1983,Su2001,Olson1999}. This method limits the diffusion coefficients so that particles diffusion velocity will not diverge. For example, the diffusion coefficient in Larsen's {\em ad hoc} flux limiter (FL) is~\cite{Olson1999}:
\begin{equation}
D(\vec{r},t)=\left[(3\sigma_{t}(T_m(\vec{r},t)))^n+\left(\frac{1}{E(\vec{r},t)}\frac{\partial{E(\vec{r},t)}}{{\partial{x}}}\right)^{n}\right]^{-\nicefrac{1}{n}}
\label{Larsen}
\end{equation}
If the gradient of $E(\vec{r},t)$ is small, the diffusion coefficient tends to the classic value of diffusion theory, $D(\vec{r},t)=1/\left[3\sigma_t(T_m(\vec{r},t))\right]$. If the gradient of $E(\vec{r},t)$ is large, Eq.~\ref{Larsen} limits the diffusion coefficient, forcing $F(\vec{r},t)\leqslant cE(\vec{r},t)$. Using $n=1$ this Flux-limiter tends to Wilson-sum FL, and taking $n\to\infty$, it tends to Wilson-Max FL~\cite{Pomraning_survey}.

There are various versions of different Flux-Limiters~\cite{Pomraning_survey,LEVERMORE1983,Su2001,Olson1999}, some of them are more physically-based than others. For example, we introduce here the well-known Levermore-Pomraning (LP)~\cite{LevermorePomraning1981,Pomraning1984}. By defining of $\omega_{\mathrm{eff}}(\vec{r},t)$, the mean number of particles emitted per collision as:
\begin{equation}
\omega_{\mathrm{eff}}(\vec{r},t)=\frac{\sigma_{s}(T_m(\vec{r},t))E(\vec{r},t)+\sigma_{a}(T_m(\vec{r},t))B(\vec{r},t)+S(\vec{r},t)/c}
{\sigma_{t}E(\vec{r},t)},
\label{omegaeff}
\end{equation}
and the normalized radiation energy density gradient $R(\vec{r},t)$ as:
\begin{equation}
R(\vec{r},t)=\frac{\vert\vec{\nabla}{E(\vec{r},t)} \vert}{\omega_\mathrm{eff}(\vec{r},t)\sigma_{t}(T_m(\vec{r},t))E(\vec{r},t)}\, 
\end{equation}
the diffusion coefficient ($D(\vec{r},t)$) in Eq.~\ref{ficks} and Eq.~\ref{diff} takes the form:
\begin{equation}
 D(\vec{r},t)=\frac{\lambda(R(\vec{r},t))}{\omega_{\mathrm{eff}}(\vec{r},t)}
\label{LPD}
\end{equation}
where $\lambda(R(\vec{r},t))$ is:
\begin{equation}
\lambda(R(\vec{r},t))=\left[\coth(R(\vec{r},t))-\frac{1}{R(\vec{r},t)}\right]\frac{1}{R(\vec{r},t)}
\label{LPD2}
\end{equation}

Another class of approximations is the variable Eddington factor (VEF) approximations. In these approximations, that have a $P_1$ notation, the second-moment term in Eq.~\ref{Rad2} is approximated with an Eddington Factor (EF), $\chi(\vec{r},t)$:
\begin{equation} 
\frac{1}{c} \frac{\partial F(\vec{r},t)}{\partial t}+ c\vec{\nabla}(\chi(\vec{r},t)E(\vec{r},t))+\sigma_{t}(T_m(\vec{r},t))F(\vec{r},t)=0,
\label{Rad2VEF}
\end{equation}
where $\chi(\vec{r},t)$ is called the Eddington factor (EF). The EF depends at $\vec{f}(\vec{r},t)$, the ratio between the first two moments:
\begin{equation} 
\vec{f}(\vec{r},t) = \frac{\vec{F}(\vec{r},t)}{cE(\vec{r},t))}.
\label{fVEF}
\end{equation}
For example, in the LP VEF~\cite{Pomraning_survey,Pomraning1982}:
\begin{equation} \vert\vec{f}(\vec{r},t)\vert=\coth(z(\vec{r},t))-1/z(\vec{r},t)
\end{equation}
and
\begin{equation} \chi(\vec{r},t)=\coth(z(\vec{r},t))[\coth(z(\vec{r},t))-1/z(\vec{r},t)].
\end{equation}
This VEF is associated with the LP Flux-limiter, (the connection is presented in~\cite{Pomraning_survey,Pomraning1982}).

\subsection{Asymptotic Diffusion and asymptotic $P_1$ (Telegrapher's Equation) Approximations}
\label{s5}
A common modified version of the diffusion approximation is the asymptotic diffusion approximation~\cite{Frankel1953,Case1953}. In this approximation, the classic Fick's law (Eq.~\ref{ficks}) is replaced by a modified (media-dependent) Fick's law, that is derived from the exact time-independent asymptotic distribution (in an infinite homogeneous medium, far away from boundaries and strong sources). In this approximation, the classic diffusion coefficient is replaced with a media ($\omega_{\mathrm{eff}}(\vec{r},t)$-dependent) diffusion coefficient:
\begin{equation}
D(\vec{r},t)=\frac{1-\omega_{\mathrm{eff}}(\vec{r},t)}{\varkappa_0^2(\vec{r},t)\sigma_{t}(T_m(\vec{r},t))}\equiv\frac{D_0(\vec{r},t)}{\sigma_{t}(T_m(\vec{r},t))}
\label{Asymptotic diffusion}
\end{equation}
$\varkappa_0(\vec{r},t)$ is the solution of the transcendental equation, which depends in $\omega_{\mathrm{eff}}(\vec{r},t)$:
\begin{equation}
 \varkappa_0(\vec{r},t)=\tanh\left(\frac{ \varkappa_0(\vec{r},t)}{\omega_{\mathrm{eff}}(\vec{r},t)}\right)
\label{kappa}
\end{equation}
The numerical values of $\varkappa_0(\omega_{\mathrm{eff}})$ and $D_0(\omega_{\mathrm{eff}})$ were tabulated extensively in~\cite{Case1953}. We note that although the asymptotic diffusion approximation produces the correct spatial asymptotic behavior, it still yields infinite particle velocities, missing the correct front (tail) behavior.

In~\cite{Heizler2010,Heizler2012}, a time-dependent analogy in a $P_1$-representation was offered, which is called the asymptotic $P_1$ approximation. In this approximation, a modified $P_1$ equation replaces the classic approximated $P_1$ equation (Eq.~\ref{Rad2Class}) with two media-dependent coefficients, $\A(\vec{r},t)$ and $\B(\vec{r},t)$:
\begin{equation} 
\frac{\A(\vec{r},t)}{c}\frac{\partial \vec{F}(\vec{r},t)}{\partial t}+c\vec{\nabla}E(\vec{r},t)+\B(\vec{r},t)\sigma_{t}(T_m(\vec{r},t))\vec{F}(\vec{r},t)=0
\label{Rad2H}
\end{equation}
$\A(\vec{r},t)$ and $\B(\vec{r},t)$ have an explicit form dependent on $\omega_{\mathrm{eff}}(\vec{r},t)$~\cite{Heizler2010,Heizler2012,Ravetto_Heizler2012}. We note that $\B(\vec{r},t)=1/D_0(\vec{r},t)$ ($D_0(\vec{r},t)$ is the asymptotic diffusion coefficient (Eq.~\ref{Asymptotic diffusion})). The full numerical expressions for $\A(\omega_{\mathrm{eff}})$ and $\B(\omega_{\mathrm{eff}})$ are described in Appendix A.

We summarize the setting:
\begin{itemize}
\item {Using the nominal $\A(\omega_{\mathrm{eff}})$ and $\B(\omega_{\mathrm{eff}})$ is called the asymptotic $P_1$ approximation ($\A\B$ approximation).}
\item {$\B(\omega_{\mathrm{eff}})=1/D_0(\omega_{\mathrm{eff}})$ (of Eq.~\ref{Asymptotic diffusion}) and $\A=0$ yields the asymptotic diffusion approximation, and hence, we will call it $\B(\vec{r},t)$  Diffusion approximation ($\B$ approximation).}
\item {$\A=\B=3$ yields the classic $P_1$ approximation.}
\item {$\B=3$ and $\A=0$ yields the classic diffusion approximation.}
\item {$\B=3$ and $\A=1$ yields the {\em ad hoc} $P_{\nicefrac{1}{3}}$ approximation~\cite{Olson1999} (In~\cite{Ravetto_Heizler2012}, we also offer the asymptotic $P_{\nicefrac{1}{3}}$ approximation, setting $\B(\omega_{\mathrm{eff}})=1/D_0(\omega_{\mathrm{eff}})$ and $\A=1$).}
\end{itemize} 
Table~\ref{table:2} summarizes all the methods itemized above. The results obtained are presented in graphs that will be discussed at a later stage of this paper.
\begin{table}
\begin{center}
\begin{tabular}{||l | l| c | l||} 
 \hline
  & Method  & In Figures  & Basic assumptions  \\[0.1ex] 
 \hline\hline
1 & IMC Simulation & \ref{fig:NL_Olson ProblemT1},\ref{fig:NL_Olson ProblemT5} &  Statistical implicit Monte Carlo approach.\\
\hline
2 & \multirow{2}{*}{$S_N$ Simulation} & \multirow{2}{*}{\ref{fig:SuOlsonBasic}, \ref{fig:SuOlsonLimiter}, \ref{fig:SuOlsonLimiterV}, \ref{fig:SuOlsonLimiterV_cs_0_5}, \ref{fig:NL_Olson ProblemT5}, \ref{fig:SuOlsonAlphBetCa1p0AB}, \ref{fig:SuOlsonAlphBetCa0p5AB}} & Solves the transport equation in $N$ \\& & & discrete ordinates.\\
\hline
3 & Classic Diffusion  & \multirow{5}{*}{\ref{fig:SuOlsonBasic}, \ref{fig:SuOlsonLimiter}, \ref{fig:SuOlsonLimiterV}, \ref{fig:SuOlsonLimiterV_cs_0_5}, \ref{fig:NL_Olson ProblemT1}, \ref{fig:NL_Olson ProblemT5}}  & The specific intensity is a sum of its only two first\\ &   & &moments ($\int_{4\pi}{ I(\hat{\Omega},\vec{r},t)\hat{\Omega}\hat{\Omega}d\Omega}\approx E(\vec{r},t)/3$),\\  & & &the derivative of the energy flux $\vec{F}(\vec{r},t)$ with \\  & && respect to time inside Eq.~\ref{Rad2Class} is negligible. \\
\hline
4 & \multirow{2}{*}{Classic $P_1$} & \multirow{2}{*}{\ref{fig:SuOlsonBasic}, \ref{fig:SuOlsonLimiter}, \ref{fig:SuOlsonLimiterV}, \ref{fig:SuOlsonLimiterV_cs_0_5}, \ref{fig:NL_Olson ProblemT1}, \ref{fig:NL_Olson ProblemT5} }  & The specific intensity is a sum of its only two
first \\& & &moments ($\int_{4\pi}{ I(\hat{\Omega},\vec{r},t)\hat{\Omega}\hat{\Omega}d\Omega}\approx E(\vec{r},t)/3$).  \\
\hline
5 &Larsen & -  & General diffusion approximation  \\
&Flux limiter & & when the diffusion coefficient is,\\& & &$D(\vec{r},t)=\left[(3\sigma_{t}(T_m(\vec{r},t)))^n+\left(\frac{1}{E(\vec{r},t)}\frac{\partial{E(\vec{r},t)}}{{\partial{x}}}\right)^{n}\right]^{-\nicefrac{1}{n}}$\\
\hline
6& LP & \ref{fig:SuOlsonLimiter}, \ref{fig:SuOlsonLimiterV}, \ref{fig:SuOlsonLimiterV_cs_0_5}   & General diffusion approximation\\&Flux limiter && when the diffusion coefficient is, $D(\vec{r},t)=\frac{\lambda(R(\vec{r},t))}{\omega_{\mathrm{eff}}(\vec{r},t)}$,\\& & & $\lambda(R(\vec{r},t))=\left[\coth(R(\vec{r},t))-\frac{1}{R(\vec{r},t)}\right]\frac{1}{R(\vec{r},t)}$,\\& & &  and $R(\vec{r},t)=\frac{\vert\vec{\nabla}{E(\vec{r},t)} \vert}{\omega_\mathrm{eff}(\vec{r},t)\sigma_{t}(T_m(\vec{r},t))E(\vec{r},t)}$  \\
\hline
7 &LP Eddington factor&  \ref{fig:SuOlsonLimiter}, \ref{fig:SuOlsonLimiterV}, \ref{fig:SuOlsonLimiterV_cs_0_5}   & General $P_1$ approximation when: \\
&  & &$\int_{4\pi}{ I(\hat{\Omega},\vec{r},t)\hat{\Omega}\hat{\Omega}d\Omega}=\vec{\nabla}(\chi(\vec{r},t)E(\vec{r},t))$ \\
& & & $\vec{f}(\vec{r},t)$, the ratio between the first two moments:
\\
& & & $\vec{f}(\vec{r},t) = \frac{\vec{F}(\vec{r},t)}{cE(\vec{r},t))}$.
\\
& & & 
 $\vert\vec{f}(\vec{r},t)\vert=\coth(z(\vec{r},t))-1/z(\vec{r},t)$\\
& & & and $ \chi(\vec{r},t)=\coth(z(\vec{r},t))[\coth(z(\vec{r},t))-1/z(\vec{r},t)]$.\\
\hline
8 &Asymptotic & \ref{fig:SuOlsonBasic}  & General diffusion approximation,  \\
&diffusion  & & $D(\vec{r},t)=\frac{1-\omega_{\mathrm{eff}}(\vec{r},t)}{\varkappa_0^2(\vec{r},t)\sigma_{t}(T_m(\vec{r},t))}\equiv\frac{D_0(\vec{r},t)}{\sigma_{t}(T_m(\vec{r},t))}$ \\
& & &
$\varkappa_0(\vec{r},t)=\tanh\left(\frac{ \varkappa_0(\vec{r},t)}{\omega_{\mathrm{eff}}(\vec{r},t)}\right)$\\
\hline
9 &Asymptotic & \ref{fig:SuOlsonBasic}  & $P_1$ approximation, in $\A\B$ form.  \\
&$P_1$& & $\B(\vec{r},t)=1/D_0(\omega_{\mathrm{eff}})$ \\
\hline
\end{tabular}
\caption{Summary of approximations discussed in Section~\ref{s2}.}
\label{table:2}
\end{center}
\end{table}

\section{The Discontinuous Asymptotic $P_1$ (Telegrapher's Equation) Approximation}
\label{s6}

The asymptotic approximations supplied good descriptions of the transport problem in isotropic media. However, in highly anisotropic media, such as
sharp boundaries or strong sources, the asymptotic solutions fail to mirror exactly how the radiation behaves. For example, solving the problem of two adjacent semi-infinite half-spaces (the two-region Milne problem)~\cite{Korn1967,mccormick1,mccormick2,mccormick3,ganapol_pomraning}, the exact solution is decomposed from an asymptotic part, which tends to the exact solution far from the boundary, and a transient part, which decays relatively fast from the boundary. Actually, this is a generalization of the classic Milne problem~\cite{Milne1921,Zeldovich2002}. Originally, Milne calculated the angular distribution of the radiated flux from a photosphere of a star. He treated the star as a semi-infinite half-space with a vacuum boundary conditions.

\begin{figure}[htbp!]
\centering 
\includegraphics*[width=8.5cm]{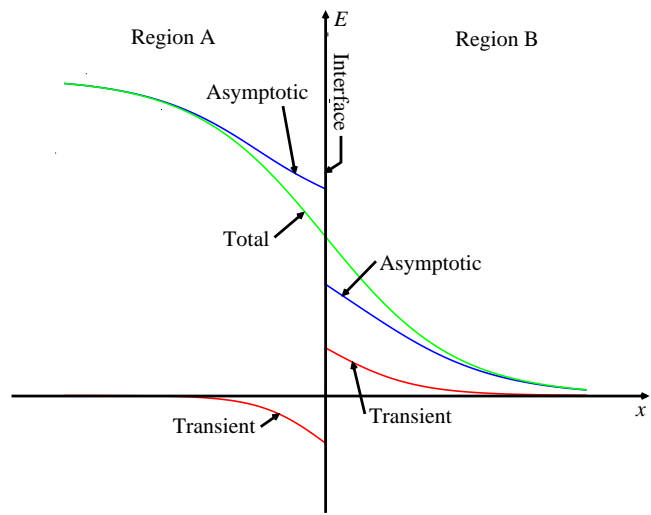}
\caption[TwoRegion]{A schematic description
of the energy density near the boundary between different regions, based on~\cite{doyas_koponen}. The asymptotic solution is discontinuous (solid blue curve) and tends to the exact solution (green) far from the boundary.
The transient part (red) is relevant near the boundary and decay relatively fast far from the boundary.}  
\label{fig:TwoRegion}
\end{figure}

In Fig.~\ref{fig:TwoRegion} we can see a schematic description of the energy density near the boundary between two different regions, based on~\cite{doyas_koponen}. Both the asymptotic (solid blue curve) and the transient part (solid red) of the solution are discontinuous, when the exact (solid green) is of course, continuous. The solution (both the asymptotic and transient parts) depends on the properties of the media, via different $\omega_{\mathrm{eff}}(\vec{r},t)$.

McCormick et. al. solved and tabulated the two-region Milne problem exactly~\cite{mccormick1,mccormick2,mccormick3}, defining the exact jump conditions of both the asymptotic scalar flux ($\rho_{\nicefrac{2}{1}}=\phi^A_{\mathrm{as}}/\phi^B_{\mathrm{as}}$) and the current density ($j_{\nicefrac{2}{1}}=J^A_{\mathrm{as}}/J^B_{\mathrm{as}}$), the first two moments, as a function of the $\omega_{\mathrm{eff}}(\vec{r},t)$ of the two media, $\omega_{\mathrm{eff}}^A$ and $\omega_{\mathrm{eff}}^B$. We note that the two-region Milne problem was solved in many other studies, for example~\cite{Korn1967,ganapol_pomraning}. McCormick et. al. used this tabulation to solve reactor-reflector problems (in a one-dimensional one-group), using a diffusion approximation with these discontinuity (jump) conditions, exactly~\cite{doyas_koponen2}.

Zimmerman~\cite{zimmerman1979} derived a simple approximation for this two-region boundary problem. In this approximation which is based on a Marshak-like approximation for the exact Milne BC for the two regions problem, the first moment (the energy flux $\vec{F}(\vec{r},t)$) is continuous, but the zero's moment (the energy density $E(\vec{r},t)$), is discontinuous. Thus, this approximation conserves particles, and is preferable for time-dependent calculations. Zimmerman expanded this method for deriving a modified discontinuous diffusion approximation. We present a short introduction to this derivation in Sec.~\ref{s6b}.

Next, in Sec.~\ref{ours} we will present our analogy for a full time-dependent $P_{1}$ asymptotic approximation. In each region, the asymptotic $P_1$ approximation is valid, and we apply the Zimmerman's discontinuous boundary condition to the energy density. We also generalize this approach for the entire space, deriving the discontinuous asymptotic $P_1$ equations.

\subsection{The Discontinuous Asymptotic Diffusion Approximation (Zimmerman's $\mu\B$ Approximation)}
\label{s6b}
Using Diffusion (or $P_1$) approximations, boundary conditions can be satisfied in an integral sense. Zimmerman used the Marshak boundary condition for the incoming flux (when vacuum is a specific case)~\cite{zimmerman1979}. In this case, the left and right boundary conditions, located in surface $\vec{r}_S$~\cite{Pomraning1973}:
\begin{subequations}
\label{partialF}
\begin{equation}
\label{partialF1}
\vec{F}_{+}(\vec{r_S},t)=
\int_{\hat\Omega\cdot\hat{n}>0}I(\hat\Omega)\hat\Omega\cdot\hat{n} d\hat\Omega=\frac{\mu(\vec{r_S},t)}{2}cE(\vec{r_S},t)+\frac{1}{2}\vec{F}(\vec{r_S},t)
\end{equation}
\begin{equation}
\label{partialF2}
\vec{F}_{-}(\vec{r_S},t)=
\int_{\hat\Omega\cdot\hat{n}<0}I(\hat\Omega)\hat\Omega\cdot\hat{n} d\hat\Omega=\frac{\mu(\vec{r_S},t)}{2}cE(\vec{r_S},t)-\frac{1}{2}\vec{F}(\vec{r_S},t)
\end{equation}
\end{subequations}
where $\hat{n}$ is the unit vector perpendicular to the surface, and:
\begin{equation}
\mu(\vec{r_S},t)=\begin{cases}
     \frac{\omega_{\mathrm{eff}}(\vec{r_S},t)}{2 \varkappa_0^2(\vec{r_S},t)}\ln\left(\frac{1}{1-\varkappa_0^2(\vec{r_S},t)}\right), & \omega_{\mathrm{eff}}(\vec{r_S},t)<1  \\
     \frac{\omega_{\mathrm{eff}}(\vec{r_S},t)}{2 \varkappa_0^2(\vec{r_S},t)}\ln\left(1+\varkappa_0^2(\vec{r_S},t)\right), &\omega_{\mathrm{eff}}(\vec{r_S},t)>1
    \end{cases}
  \label{muAsymptotic}
\end{equation}
The spatial and temporal dependence of $\mu(\vec{r},t)$ is due to $\omega_{\mathrm{eff}}$, as it is for $\A(\vec{r},t)$, and $\B(\vec{r},t)$. The full expression of $\mu(\omega_{\mathrm{eff}})$ is in Appendix A.

Looking at a boundary between two different media (Fig.~\ref{fig:TwoRegion}), the flux comes out of medium A, $\vec{F}_{-}^A(\vec{r_S},t)$, is the incoming flux of medium B, $\vec{F}_{+}^B(\vec{r_S},t)$, and vice versa: 
\begin{subequations}
\label{fpm}
\begin{equation}
\vec{F}_{+}^A(\vec{r_S},t)=\vec{F}_{-}^B(\vec{r_S},t)
\label{f1}
\end{equation}
\begin{equation}
\vec{F}_{+}^B(\vec{r_S},t)=\vec{F}_{-}^A(\vec{r_S},t)
\label{f2}
\end{equation}
\end{subequations}
Adding and subtracting Eqs.~\ref{fpm}, and using the definitions of Eqs.~\ref{partialF} yield continuous flux ($\vec{F}(\vec{r_S},t)$), and thus energy conservation), and a discontinuity in the energy density ($E(\vec{r_S},t)$):
\begin{subequations}
\begin{equation}
\vec{F}_A(\vec{r_S},t)=\vec{F}_B(\vec{r_S},t)
\label{dis_f2}
\end{equation}
\begin{equation}
\mu_AE_A(\vec{r_S},t)=\mu_BE_B(\vec{r_S},t)
\label{dis_f1}
\end{equation}
\label{dis_f}
\end{subequations}

It can be shown that (assuming the asymptotic diffusion theory is valid far from the boundary) Eqs.~\ref{dis_f} yields a modified discontinuous Fick's law~\cite{zimmerman1979}:
\begin{equation}
\vec{F}(\vec{r},t)=-\frac{cD(\vec{r},t)}{\mu(\vec{r},t)}\vec{\nabla}\left(\mu(\vec{r},t)E(\vec{r},t)\right),
\label{ZFick}
\end{equation}
i.e., Zimmerman extended the discontinuity jump conditions, for an entire non-uniform space.
Substituting Eq.~\ref{ZFick} in the conservation law, Eq.~\ref{Rad1} yields a new discontinuous asymptotic diffusion approximation:
\begin{equation}
\\ \frac{1}{c}\frac{\partial E(\vec{r},t)}{\partial t}-
\\ \vec{\nabla}\left(\frac{D(\vec{r},t)}{\mu(\vec{r},t)}\vec{\nabla}\left(\mu(\vec{r},t){E}(\vec{r},t)\right)\right)=
\\ \sigma_{a}((T_m(\vec{r},t))\left(\frac{B(\vec{r},t)}{c}-E(\vec{r},t)\right)+\frac{S(\vec{r},t)}{c}
\label{Zdiff}
\end{equation}
Since Eqs.~\ref{ZFick} and~\ref{Zdiff} contain two medium-dependent variables, $\mu(\omega_{\mathrm{eff}})$ and $D_0(\omega_{\mathrm{eff}})$, we call it the $\mu\B$ approximation (recalling that $\B(\omega_{\mathrm{eff}})=1/D_0(\omega_{\mathrm{eff}})$, see Sec.~\ref{s5}).

We note that there are similar works~\cite{Pomraning1965,Pomraning_Nukleonik}, deriving similar discontinuous Fick's law (using $\beta(\omega_{\mathrm{eff}})$ as the discontinuity in the energy density and continuous flux). These works produce, from a different point of view, values close to Zimmerman's $\mu(\omega_{\mathrm{eff}})$. In addition, a discontinuous Fick's law based on the $P_2$ approximation yields also good results in some neutronics problems~\cite{rulko_larsen}.

\subsection{Derivation of the Discontinuous Asymptotic $P_1$
Approximation ($\mu\A\B$ Approximation)}
\label{ours}

Using the discontinuity jump conditions from the previous section, we can derive a time-dependent analogy, now in a full $P_1$ form (instead of a Fick's law form in the time-independent case). This approximation contains both $\A(\omega_{\mathrm{eff}})$ and $\B(\omega_{\mathrm{eff}})$ from the asymptotic $P_1$ approximation, and the jump condition variable $\mu(\omega_{\mathrm{eff}})$, yielding the Discontinuous Asymptotic $P_1$ Approximation (or in short, the $\mu\A\B$ Approximation).

First, in each region (see Fig.~\ref{fig:TwoRegion}) the asymptotic $P_1$ equations are valid, Eqs.~\ref{Rad2H} and~\ref{Rad1}. Suppose that the boundary is located in the origin, i.e. $\vec{r}_S=0$, we can rewrite Eq.~\ref{Rad2H} from the two sides of the origin:
\begin{subequations}
\label{our1}
\begin{align}
\label{our1a}
 & c\frac{E(\Delta \vec{r},t)-E(\vec{r},t)\vert_{\vec{r}\to0+}}{\Delta \vec{r}}=-\frac{\A(\Delta \vec{r},t)}{c}\frac{\partial \vec{F}(\vec{r},t)}{\partial t}\biggr|_{\vec{r}\to 0+}- \nonumber \\
 & \B(\Delta \vec{r},t){\sigma_{t}((T_m(\Delta\vec{r},t))}\vec{F}(\vec{r},t)\vert_{\vec{r}\to0+}
\end{align}
\begin{align}
\label{our1b}
& c\frac{E(-\Delta  \vec{r},t)-E(\vec{r},t)\vert_{\vec{r}\to0-}}{\Delta  \vec{r}}=\frac{\A(-\Delta \vec{r},t)}{c}\frac{\partial \vec{F}(\vec{r},t)}{\partial t}\biggr|_{\vec{r}\to 0-}+ \nonumber \\
& \B(-\Delta \vec{r},t){\sigma_{t}((T_m(-\Delta\vec{r},t))}\vec{F}(\vec{r},t)\vert_{\vec{r}\to0-}
\end{align}
\end{subequations}
where $E(\vec{r},t)\vert_{\vec{r}\to0+}=E_B(0,t)$ and $E(\vec{r},t)\vert_{\vec{r}\to0-}=E_A(0,t)$. $\vec{F}(\vec{r},t)\vert_{\vec{r}\to0+}=F_B(0,t)$, and $\vec{F}(\vec{r},t)\vert_{\vec{r}\to0-}=F_A(0,t)$, and their derivatives with respect to time, respectively. Multiplying Eq.~\ref{our1a} by $\mu(\Delta \vec{r},t)=\mu_B(0,t)$ and Eq.~\ref{our1b} by $\mu(-\Delta \vec{r},t)=\mu_A(0,t)$, and solving for $E(\vec{r},t)\vert_{\vec{r}\to0+}$ and $E(\vec{r},t)\vert_{\vec{r}\to0-}$ yields:
\begin{subequations}
\label{our2}
\begin{align}
& c\mu(\Delta \vec{r},t)\frac{E(\vec{r},t)\vert_{\vec{r}\to0+}}{\Delta \vec{r}}=c\mu(\Delta \vec{r},t)\frac{E(\Delta \vec{r},t)}{\Delta \vec{r}}+\mu(\Delta \vec{r},t)\A(\Delta \vec{r},t)\frac{\partial \vec{F}(\vec{r},t)}{c\partial t}\biggr|_{\vec{r}\to 0+}+ \nonumber \\
&\mu(\Delta \vec{r},t)\B(\Delta \vec{r},t){\sigma_t((T_m(\Delta\vec{r},t))}\vec{F}(\vec{r},t)\vert_{\vec{r}\to0+}
\end{align}
\begin{align}
& c\mu(-\Delta \vec{r},t) \frac{E(\vec{r},t)\vert_{\vec{r}\to0-}}{\Delta \vec{r}}=c\mu(-\Delta \vec{r},t)\frac{E(-\Delta \vec{r},t)}{\Delta \vec{r}}-\mu(-\Delta \vec{r},t)\A(-\Delta \vec{r},t)\frac{\partial \vec{F}(\vec{r},t)}{c\partial t}\biggr|_{\vec{r}\to 0-}- \nonumber \\
& \mu(-\Delta \vec{r},t)\B(-\Delta \vec{r},t){\sigma_t((T_m(-\Delta\vec{r},t))}\vec{F}(\vec{r},t)\vert_{\vec{r}\to0-}
\end{align}
\end{subequations}
Applying the discontinuity condition in $E(\vec{r},t)$, Eq.~\ref{dis_f}(a) and the conservation of flux, Eq.~\ref{dis_f}(b), and subtracting  Eqs.~\ref{our2} yields: 
\begin{align}
& c\frac{\mu(\Delta \vec{r},t)E(\Delta \vec{r},t)-\mu(-\Delta \vec{r},t) E(-\Delta \vec{r},t)}{2\Delta \vec{r}}+\frac{\A(\Delta \vec{r},t)\mu(\Delta \vec{r},t)+\A(-\Delta \vec{r},t)\mu(-\Delta \vec{r},t)}{2}\frac{\partial F(\vec{r},t)}{c\partial t}\biggr|_{\vec{r}\to 0}+ \nonumber \\
& \frac{\mu(\Delta \vec{r},t)\B(\Delta \vec{r},t)\sigma_t((T_m(\Delta\vec{r},t))+\mu(-\Delta \vec{r},t)\B(-\Delta \vec{r},t)\sigma_t((T_m(-\Delta\vec{r},t))}{2}\vec{F}(\vec{r},t)\vert_{\vec{r}\to0}=0
\end{align}
where $\vec{F}(\vec{r},t)\vert_{\vec{r}\to0}=\vec{F}(\vec{r},t)\vert_{\vec{r}\to0+}=\vec{F}(\vec{r},t)\vert_{\vec{r}\to0-}$ and $\frac{\partial F(\vec{r},t)}{\partial t}\biggr|_{\vec{r}\to 0}=\frac{\partial F(\vec{r},t)}{\partial t}\biggr|_{\vec{r}\to 0+}=\frac{\partial F(\vec{r},t)}{\partial t}\biggr|_{\vec{r}\to 0-}$ from Eq.~\ref{dis_f2}, of course.
Taking $\Delta\vec{r}\to0$ yields a general discontinuous asymptotic $P_1$ equation (for the entire space):
\begin{equation}  
\mu(\vec{r},t)\frac{\A(\vec{r},t)}{c}\frac{\partial F(\vec{r},t)}{\partial t}+c\vec{\nabla}\left({\mu(\vec{r},t)}E(\vec{r},t)\right)+
\mu(\vec{r,t})\B(\vec{r},t){\sigma_{t}((T_m(\vec{r},t))}F(\vec{r},t)=0
\label{DisC2}
\end {equation}

Eqs.~\ref{Rad1} and~\ref{DisC2} define the new approximation, the {\em discontinuous asymptotic $P_1$ approximation}. 
These equations contain three medium-dependent variables, $\mu(\omega_{\mathrm{eff}})$ and $\A(\omega_{\mathrm{eff}})$ and $\B(\omega_{\mathrm{eff}})$, and thus we call it also the $\mu\A\B$ approximation. Our new approximation has the advantage of the $P_1$ notation along with the using of the asymptotic exact solutions. It is also important to note the method reserves energy which is important for the physical meaning. 

The discontinuous asymptotic $P_1$ approximation is valid also for neutronics, replacing $E(\vec{r},t)$ and $\vec{F}(\vec{r},t)$ with $\phi(\vec{r},t)$ and $\vec{J}(\vec{r},t)$ and $\omega_{\mathrm{eff}}$ with $c$ (do not confuse with the speed of light). For a more detailed discussion, see Appendix B. Also, the extension to multi-group is straightforward due to the energy dependent definition of $\omega_{\mathrm{eff}}$ (or $c$, in the case of neutronics)~\cite{Winslow1968,Pomraning1984}.

By assuming that both the time derivative of $\vec{F}(\vec{r},t)$ and the spatial derivative of $\A/\B\sigma_t$ are small enough, we can neglect $\vec{\nabla}\left(\frac{\A(\vec{r},t)}{\B(\vec{r},t)\sigma_t(T_m(\vec{r},t))}\right)\cdot\frac{\partial \vec{F}(\vec{r},t)}{\partial t}$ and obtained from Eqs.~\ref{DisC2} and~\ref{Rad1}:
\begin{align}
& \frac{\A(\vec{r},t)}{\B(\vec{r},t)c\sigma_t}\frac{\partial^2 E(\vec{r},t)}{\partial t^2}-
\vec{\nabla}\left[\frac{c}{\sigma_{t}\mu(\vec{r},t)\B(\vec{r},t)}
\vec{\nabla}\left({\mu(\vec{r},t)}E(\vec{r},t)\right)\right]+
\frac{\partial E(\vec{r},t)}{\partial t}= \nonumber \\
& \frac{4\pi\A(\vec{r},t)}{c\B(\vec{r},t)\sigma_t}\frac{\partial(\sigma_a B(\vec{r},t))}{\partial t}-
\frac{\A(\vec{r},t)}{\B(\vec{r},t)\sigma_t}\frac{\partial(\sigma_a E)}{\partial t}+ \nonumber \\ 
& \frac{\A(\vec{r},t) }{\B(\vec{r},t)\sigma_t}\frac{\partial S(\vec{r},t)}{\partial t}+
\sigma_{a}\left({4\pi}{B(\vec{r},t)}-c E(\vec{r},t)\right)+
S(\vec{r},t)
 \label{dealing5}
\end{align}
This is the 
{\em discontinuous asymptotic Telegrapher's equation} which is our new modification of Eq.~\ref{telegraphers}.

\section{Results}
\label{results}
In this section we test the new discontinuous asymptotic $P_1$ approximation ($\mu\A\B$ approximation) numerically, with some well-known radiative transfer benchmarks. The numerical results are compared to exact benchmarks' solutions, as well as other approximations that were introduced in Sec.~\ref{s2}. The first benchmark is the well-known constant opacity Su-Olson benchmark~\cite{SuOlson1996}; the other is a variable non-linear opacity Olson's benchmark~\cite{Olson1999}. We will see that the new method seems to be more accurate than other methods,
while still being easy to apply.

\subsection{The Constant Opacity (Su-Olson) problem}

The well-known Su-Olson benchmark~\cite{SuOlson1996} is a basic non-equilibrium slab-geometry radiative transfer benchmark that uses a constant opacity in an infinite,
isotropic scattering medium. The radiation source in the medium is isotropic
and constant for a limited period (and is zero afterwards) and the material is initially cold and homogeneous. In this benchmark it is convenient to set dimensionless position $z$ and time $\tau$, and normalized radiation and material energy densities, $W$ and $V$, respectively:
\begin{equation}
x=\sigma_{t} z; \quad
\tau=\epsilon c \sigma_{t} t; \quad
V=\left(\frac{T}{T_H}\right)^4;\quad
W=\int_{-1}^{1}d\mu\frac{I(\mu)}{aT_H^4}; \quad
\label{SuOlsonUnit}
\end{equation}
$T_H$ is defined as the Hohlraum temperature (or any other reference temperature).
The material heat capacity is defined as:
$C_v=\alpha T^3$ and $\epsilon=4a/\alpha$. It is also convenient to define the ratio of the scattering cross section to the total cross section $c_s=\sigma_s/ \sigma_{t}$, since we use dimensionless position variable.
This problem has an exact solution~\cite{SuOlson1996} for a specific source term $S(x,\tau)$:
 \begin{equation}
S(x,\tau)=\begin{cases}
      1, & \text{if}\ \tau\leq 10, \quad x\leq 0.5 \\
      0, & \text{otherwise}
    \end{cases}
    \label{source}
\end{equation}

\begin{figure}[htbp!]
\centering 
(a)
\includegraphics*[width=7.5cm]{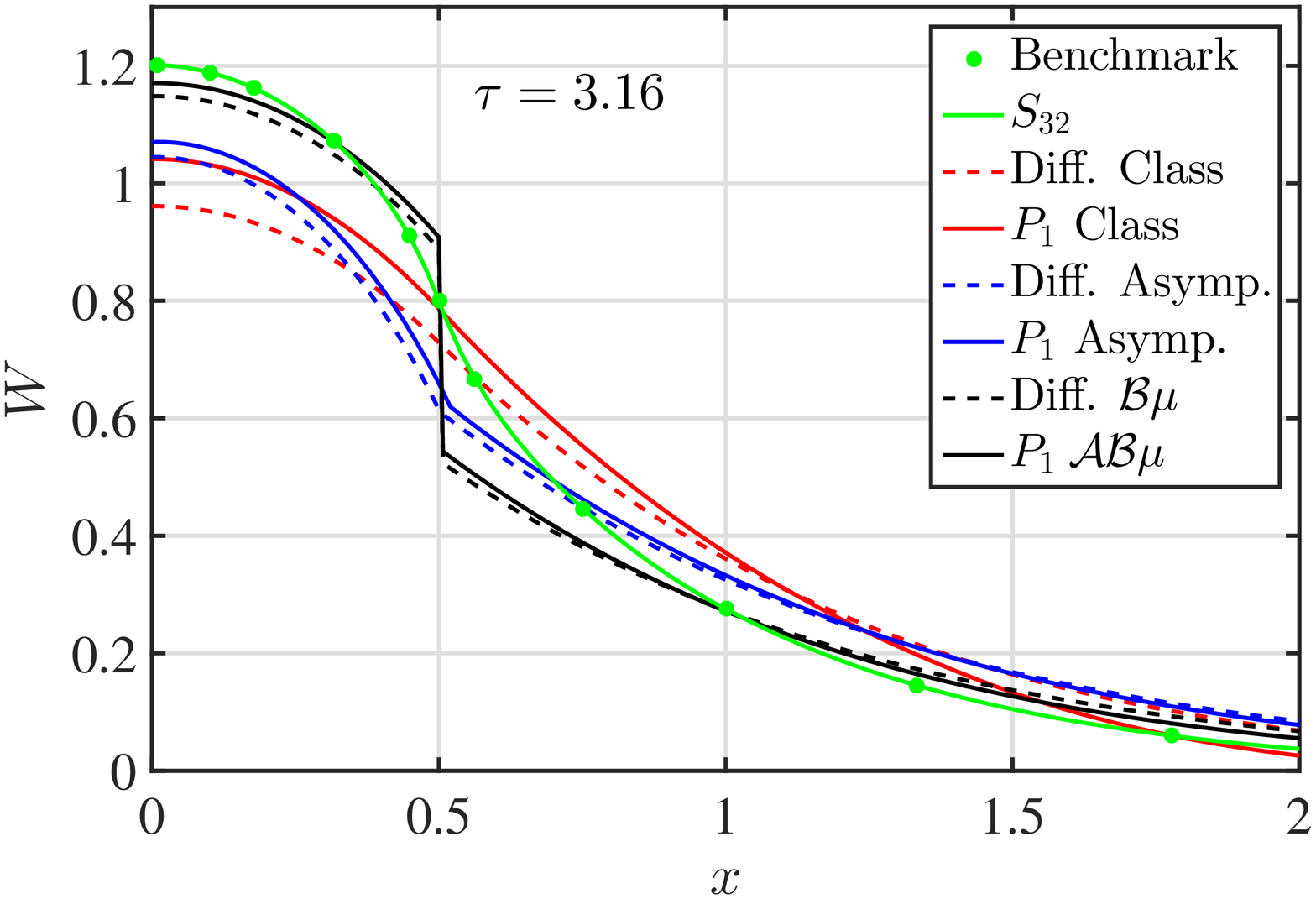}
(b)
\includegraphics*[width=7.5cm]{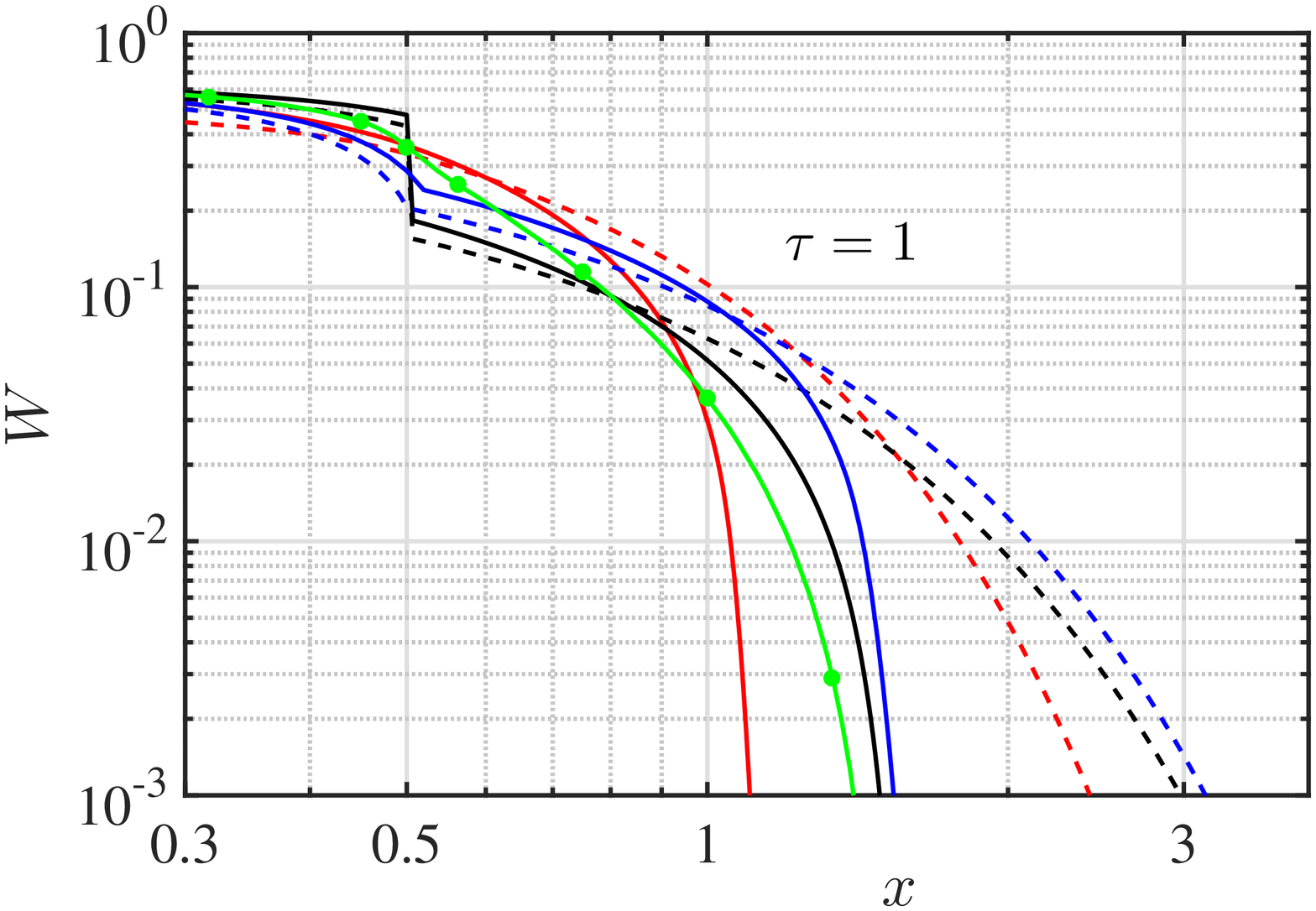}
\caption[SuOlson Problem basic]{The radiation energy density (W) in linear (a) and logarithmic (b) scales as a function
of the optical depth. The Su-Olson problem here is for a non scattering case, $c_s=0$. 
The circles are the exact transport solution which is taken from~\cite{SuOlson1996}, the green curves are the $S_{32}$. The red dashed and solid curves are the classic diffusion and $P_1$ approximations, respectively. 
The blue dashed and solid curves are the asymptotic diffusion and $P_1$ approximations.
The black dashed and solid curves are Zimmerman's $\B\mu$ diffusion and the $\A\B\mu$ $P_1$ approximations.}  
\label{fig:SuOlsonBasic}
\end{figure}

The radiation energy as a function of space is presented in Fig.~\ref{fig:SuOlsonBasic} using several approximations and the exact solution for the no scattering case, $c_s=0$.
In Fig.~\ref{fig:SuOlsonBasic}(a) the radiation energy is shown in linear scale for $\tau=3.16$,
and in Fig.~\ref{fig:SuOlsonBasic}(b) in logarithmic scale for $\tau=1$. We note that for the non-scattering case ($c_s=0$), there is an analytic solution for the classic $P_1$ approximation~\cite{McClarren2008}, and our numerical results reproduce this analytic solution.

First, the benchmark results (full symbols) and $S_{32}$ numerical solutions (green solid curves)
fit perfectly. Next, both the classic diffusion and $P_1$ approximations (dashed and solid curves) yield bulk energy results that are too low. (Fig.~\ref{fig:SuOlsonBasic}(a)). In addition, in the logarithmic scale (Fig.~\ref{fig:SuOlsonBasic}(b)) it is noticeable that the diffusion approximation heat front is too fast, while $P_1$ heat front is too slow.
The asymptotic diffusion approximation (blue dash curves) suffers from the same problems, yielding just a little bit better results than the classic diffusion approximation. The front of the asymptotic $P_1$ (blue solid curves), is quite good but has too small bulk energy, and is similar to the classic $P_1$ approximation. Zimmerman's discontinuous asymptotic diffusion approximation (the $\mu \B$ approximation), yields better results in the bulk, resulting the discontinuity jump condition, but the front is still too fast, as any diffusion approximation (because of the infinite velocity). However, it is clear that the new discontinuous asymptotic $P_1$ approximation (the $\mu\A\B$ approximation) is very close to the exact solution, both in the bulk and the front (except the jump itself). 

\begin{figure}[htbp!]
\centering 
\includegraphics*[width=7.5cm]{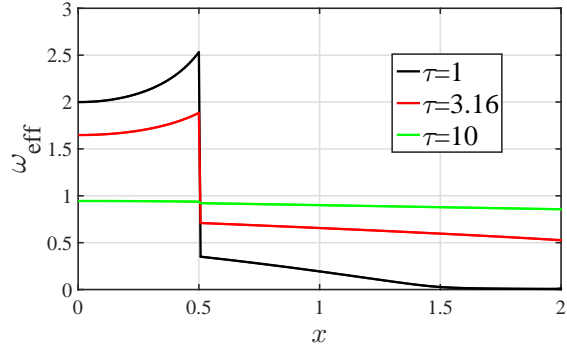}
\caption[Omega basic]{The $\omega_\mathrm{eff}$ for our $\A\B\mu$ approximation. 
Here $\omega_\mathrm{eff}$ is for three different times, $\tau=$1, 3.16, 10. The jump in $x=0.5$ is due to the step function in the source term.}
\label{fig:SuOlsonOmega}
\end{figure}
Of course, in the interface of the source (in $x=0.5$), there is a large discontinuity in the energy density (both employ the new approximation or Zimmerman's approximation). This is due to the functional dependence of $\mu(x,\tau)$ on $\omega_{\mathrm{eff}}(x,\tau)$ (Eq.~\ref{muAsymptotic}). which is a function of time and space. In Fig.~\ref{fig:SuOlsonOmega} we can see
$\omega_{\mathrm{eff}}(x,\tau)$ as a function of $x$ for several times.
The clear jump in $x=0.5$ is due to the step function of $S(x,\tau)$ (Eq.~\ref{source}), and it is mostly important in early times. 
As the energy increases in later times,
$S(x,\tau)$ is less important in the $\omega_{\mathrm{eff}}$, and the discontinuity is less apparent. 

\begin{figure}[htbp!]
\centering 
(a)
\includegraphics*[width=7.5cm]{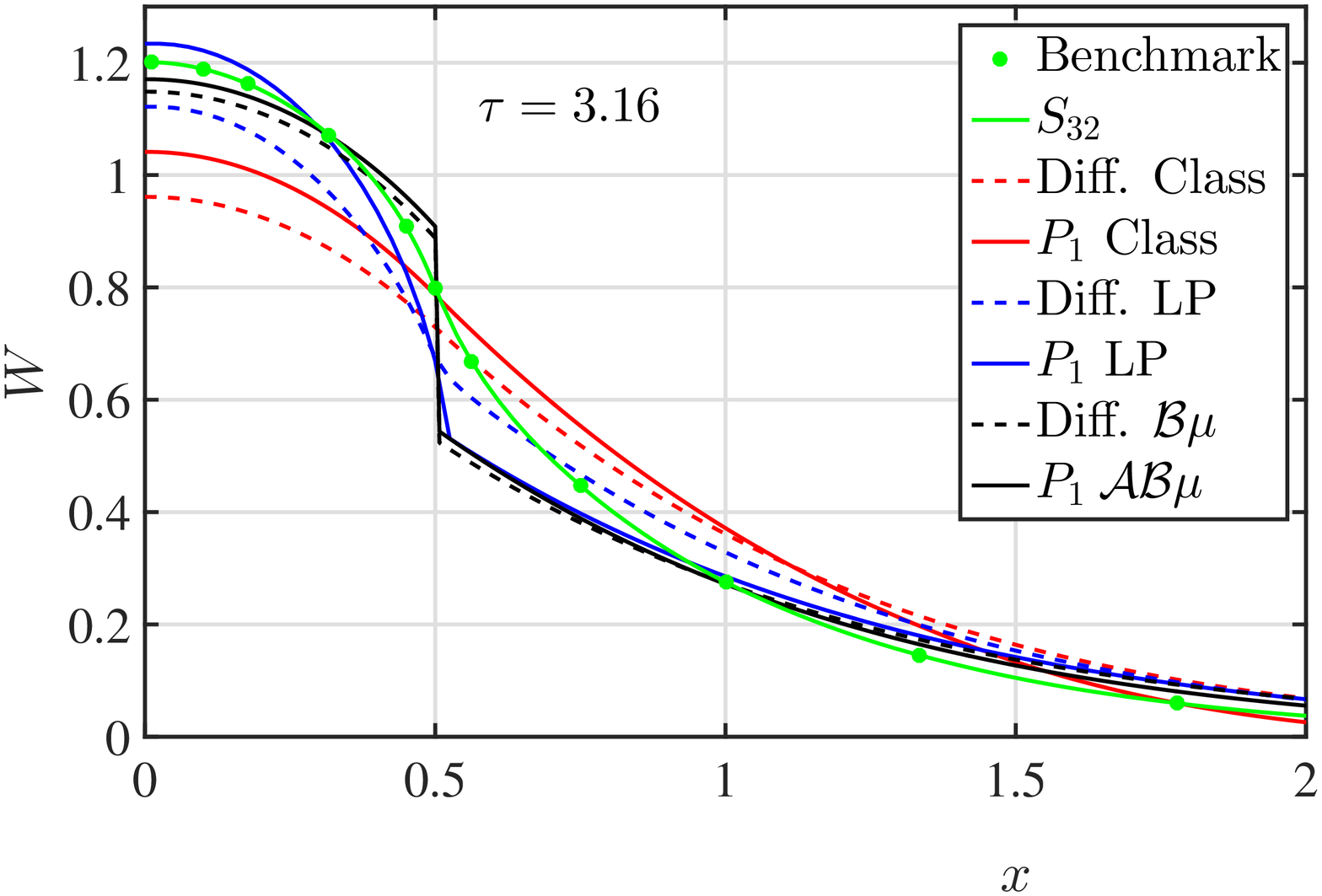}
(b)
\includegraphics*[width=7.5cm]{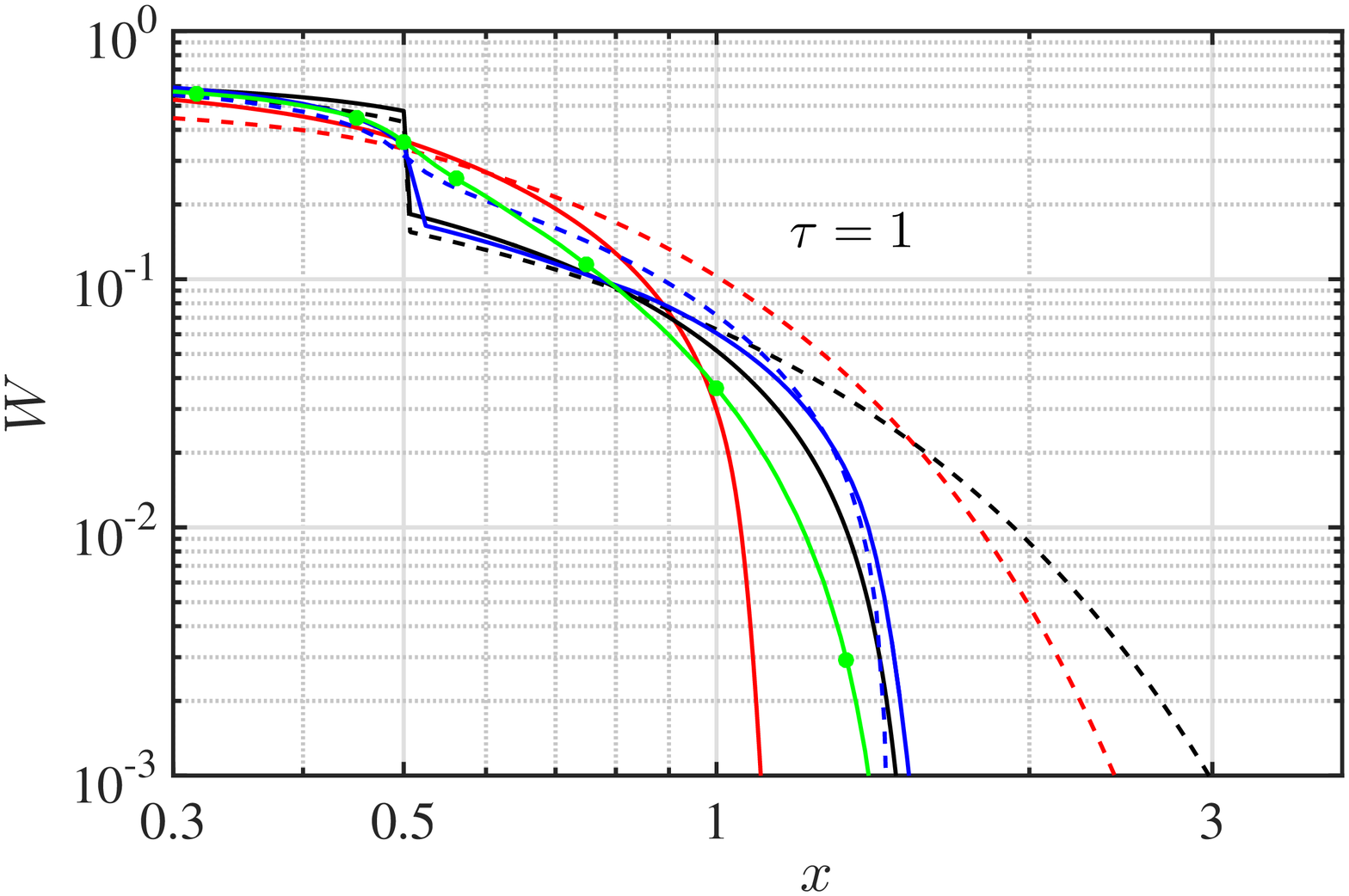}
\caption[SuOlson Problem limiter]{The Su-Olson benchmark radiation energy ($W$) with $c_s=0$ in linear scale (a) and in logarithmic scale (b). In addition to the exact results, classic diffusion and $P_1$ approximations and the discontinuous approximations as in Fig.~\ref{fig:SuOlsonBasic}, the Levermore-Pomraning FL (dashed blue curves) and the Levermore-Pomraning VEF (solid blue curves) are presented.}  
\label{fig:SuOlsonLimiter}
\end{figure}
Moreover, the new $\mu\A\B$ approximation yields better results than the gradient-dependent approximations, such as the different Flux-Limiters and variable Eddington factors approximations. In Fig.~\ref{fig:SuOlsonLimiter} (blue dashed and solid curves) we introduce the results of the Levermore-Pomraning flux limiter and Eddington factor. We found that it yields better or similar results than other flux-limiters or Eddington factors, such as Minerbo's or Kershaw's (see also in~\cite{Su2001,Olson1999}). The LP FL results are quite similar to the LP VEF results, when the latter yields slightly better results. We can see that the new $\mu\A\B$ approximation yields better results than these gradient-dependent approximations. This is extremely important since the gradient-dependent approximations are harder to apply in multi-dimensions (especially in curvilinear geometries), while the new approximation is easy to apply as a simple $P_1$ implementation.

\begin{figure}[htbp!]
\centering 
(a)
\includegraphics*[width=7.5cm]{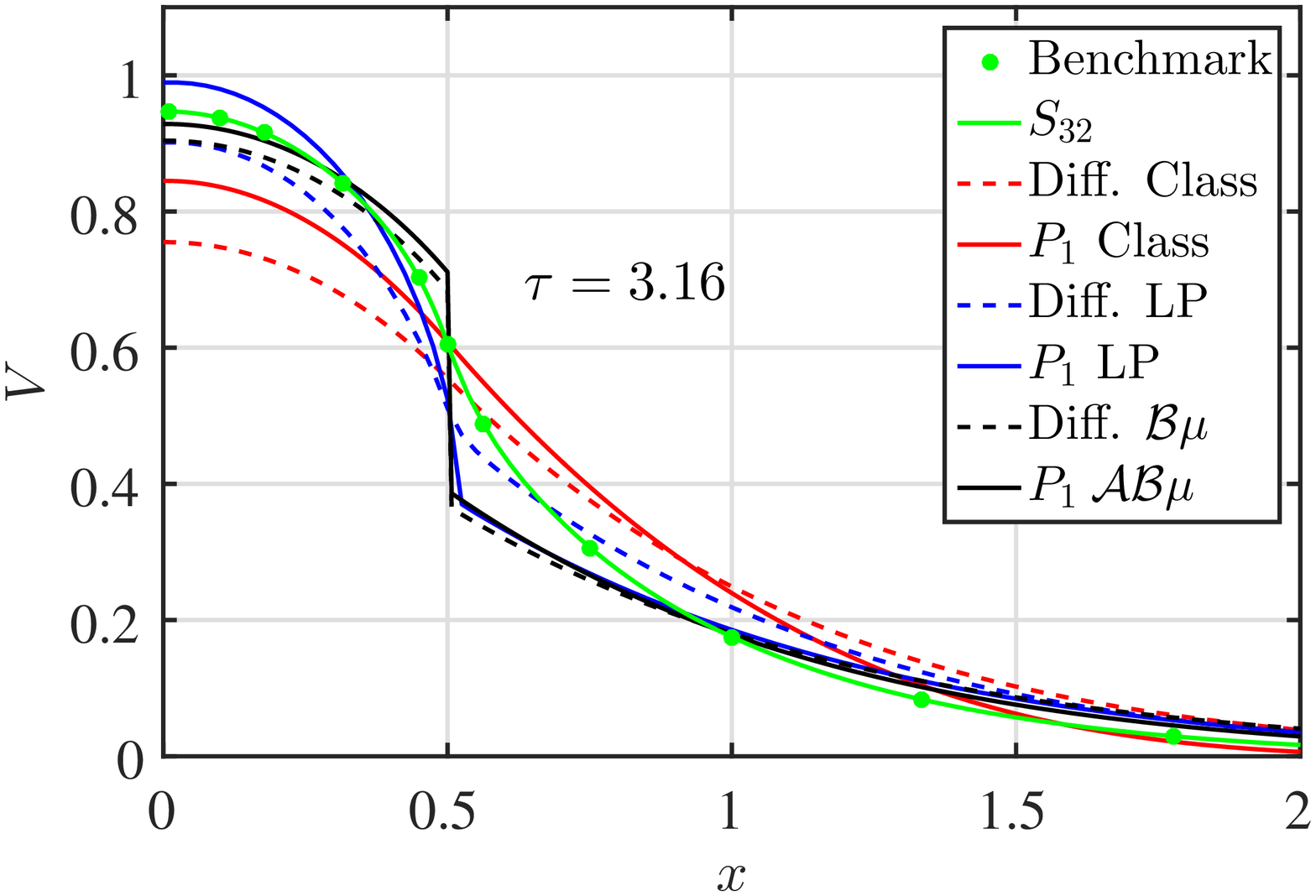}
(b)
\includegraphics*[width=7.5cm]{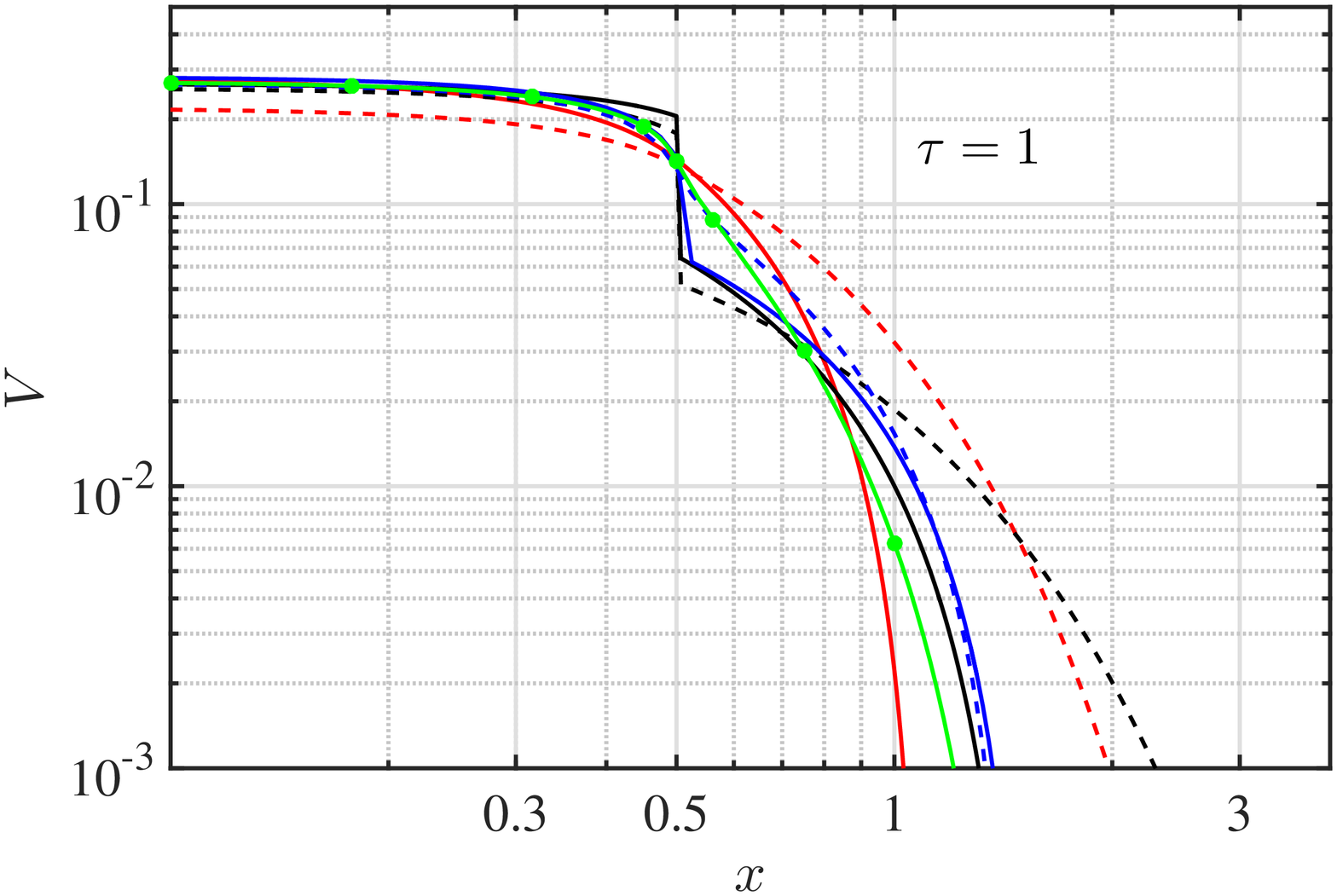}
\caption{The same as Fig.~\ref{fig:SuOlsonLimiter} but for the material energy $V$.}  
\label{fig:SuOlsonLimiterV}
\end{figure}

\begin{figure}[htbp!]
\centering 
(a)
\includegraphics*[width=7.5cm]{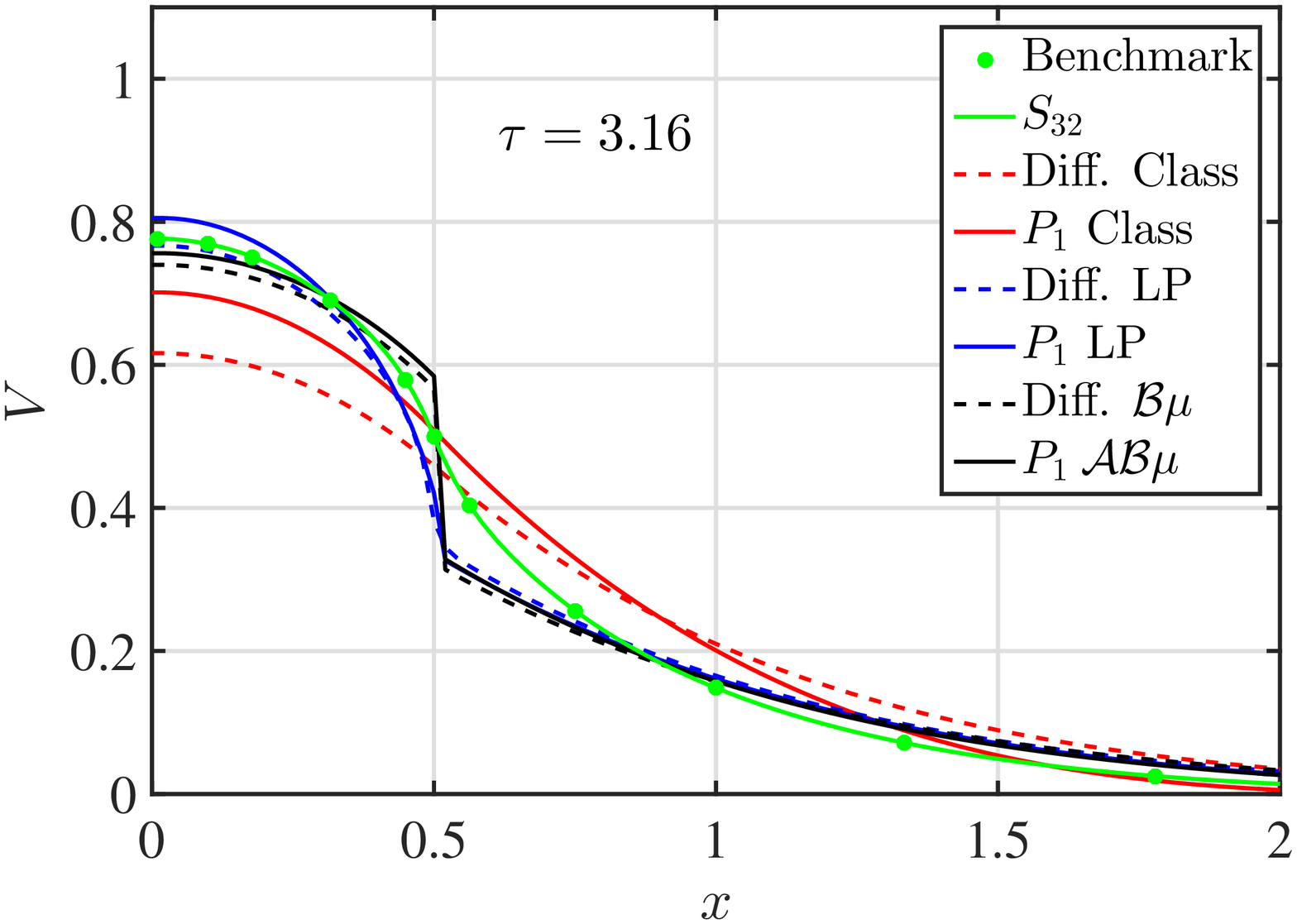}
(b)
\includegraphics*[width=7.5cm]{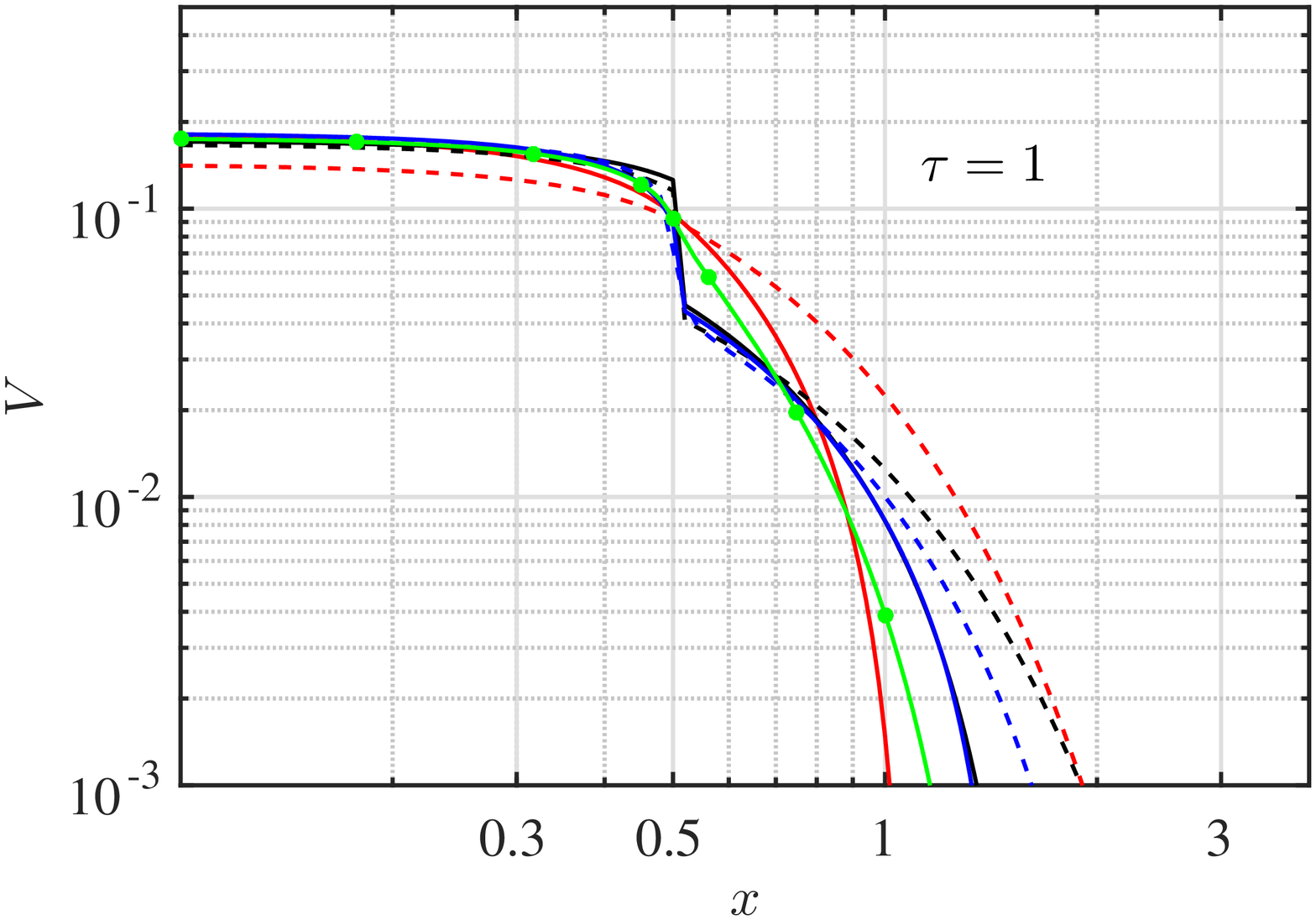}
\caption{The same as Fig.~\ref{fig:SuOlsonLimiterV} but for the scattering-included case, $c_s=0.5$.}  
\label{fig:SuOlsonLimiterV_cs_0_5}
\end{figure}
In Fig.~\ref{fig:SuOlsonLimiterV} we can see the material energy for the case of $c_s=0$. We can see that the same conclusions that were presented regarding the radiation energy, are also valid for the material energy. The new $\mu\A\B$ approximation yields the best estimations compareed to the exact results (except the jump itself, that is of course, non-physical). In Fig.~\ref{fig:SuOlsonLimiterV_cs_0_5} we can see that the same is also valid for scattering media with $c_s=0.5$ as well (we present here the material energy since the radiation energy is very close to the $c_s=0$ case, Fig.~\ref{fig:SuOlsonLimiter}). The discontinuity jump in the $c_s=0.5$ case is smaller than in the $c_s=0$ case, due to smaller differences in $\omega_{\mathrm{eff}}$ in the scattering-included case.

\subsection{Olson's non-linear opacity problem}

The assumption of constant opacity which allows the semi-analytic solution that is made in the Su-Olson
is usually, not realistic, since the opacity is a strong function of the material temperature. Therefore, Olson~\cite{Olson1999} set another benchmark, where the opacity varies with the material temperature:
\begin{equation}
\sigma_a = T^{-3}
\label{NLOlsonOpacity}
\end{equation}
In this problem, $C_v$ is constant and the dimensionless time $\tau$ is:
\begin{equation}
\tau = \frac{4acT_{H}^3}{C_v}t
\label{NLOlsontauu}
\end{equation}
We note that the $T^{-3}$ dependence is quite realistic opacity for low-Z materials such as Aluminum~\cite{Murakami1990}. Instead of an internal source term (like in the Su-Olson benchmark), Olson et. al. apply an isotropic incident radiation flux located on the slab's surface at $x=0$:
\begin{equation}
F_{in} = \frac{acT_{H}^4}{4}
\label{NLOlsonFin}
\end{equation}
Applying the Marshak boundary condition and solving for the net flux $F(0,\tau)$~\cite{Olson1999}:
\begin{equation}
F(x=0,\tau)=\begin{cases}
\mathrm{Classic} \quad P_1\mathrm{/Diffusion},  & 2F_{in} - cW(0,\tau)/2
\\ \mathrm{Asymptotic} \quad P_1\mathrm{/Diffusion},  & 2F_{in} - \mu(0,\tau) cW(0,\tau)
\end{cases}
\label{NLOlsonFin2}
\end{equation}
when $\mu(0,\tau)$ is a function of $\omega_{\mathrm{eff}}(0,\tau)$ as defined by Eq.~\ref{muAsymptotic}, assuming the asymptotic flux distribution instead of the classic $P_1$ notation~\cite{Pomraning1973}.

\begin{figure}[htbp!]
\centering 
(a)
\includegraphics*[width=7.5cm]{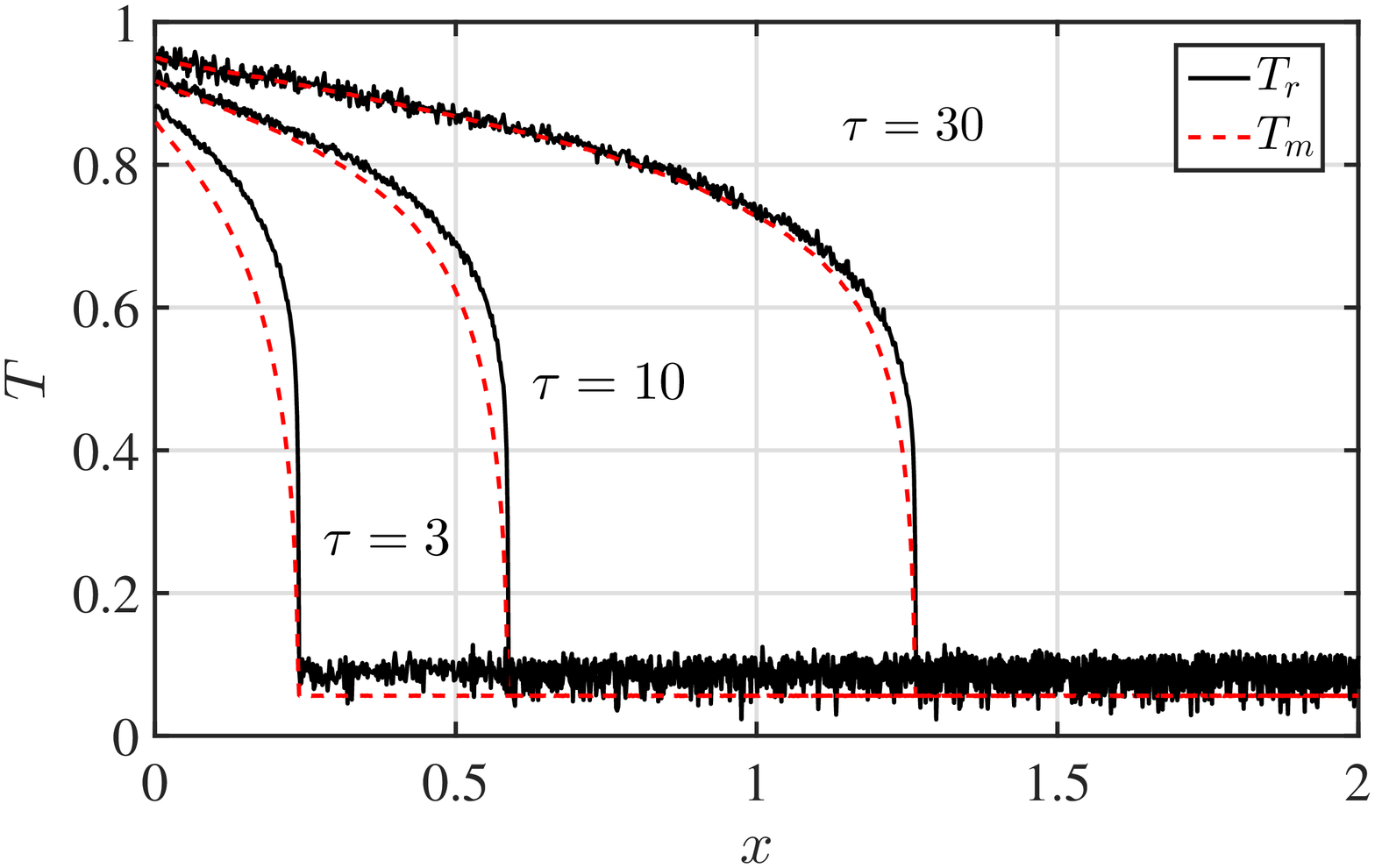}
(b)
\includegraphics*[width=7.5cm]{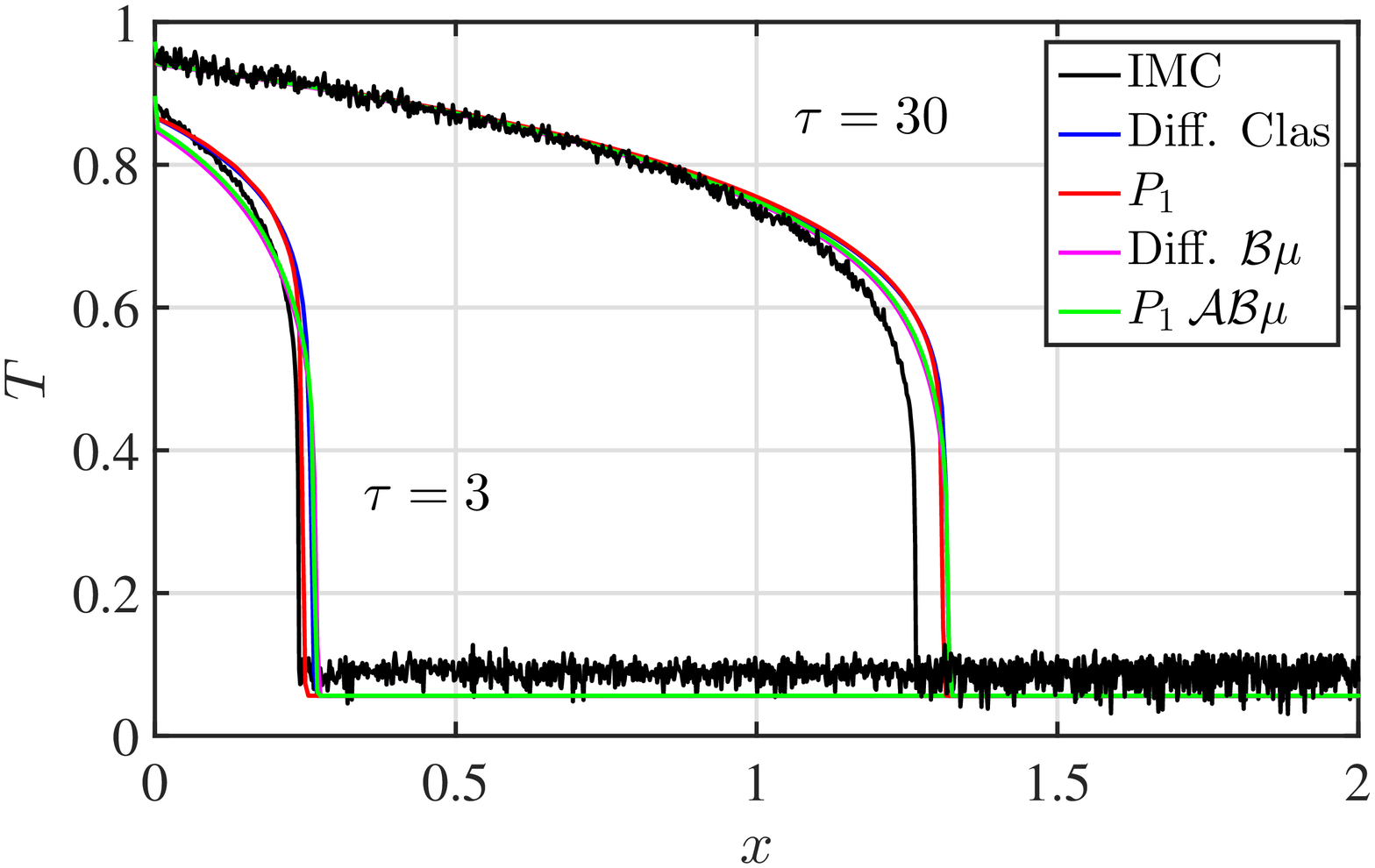}
\caption[NL_Olson ProblemT1]{(a) The radiation and material temperatures of the IMC in different times ($\tau=$3, 10, 30) for the Olson's nonlinear opacity problem (using $T_H=1$). (b) The radiation temperature ($T_r$), as a function of the optical depth in different times ($\tau=$3, 30). 
The red and blue solid curves are the classic diffusion and $P_1$. The magenta and the green curves are the $\B\mu$ diffusion and the $\A\B\mu$ $P_1$ approximations. The exact IMC is in the black solid curves.} 
\label{fig:NL_Olson ProblemT1}
\end{figure}

First we solve this problem with two exact approximations, both $S_{N}$ with $N=32$ and Implicit Monte Carlo (IMC)~\cite{IMC}.
Both methods yield precisely the same solution, so we choose to introduce explicitly here the IMC results. The results of the Olson's nonlinear opacity benchmark are shown in Fig.~\ref{fig:NL_Olson ProblemT1}. In Fig.~\ref{fig:NL_Olson ProblemT1}(a) we introduce the difference between the radiation and material temperatures. The results (of both $S_N$ and IMC) are very similar to the exact VEF that was introduced in~\cite{Olson1999}. Since in this benchmark $T_H=1$, the problem turned out to be relatively thick in optical terms, when there exists a large number of mean free paths even at early times. That is why the material temperature ($T_m$) is very close to the radiation temperature ($T_r$). In Fig.~\ref{fig:NL_Olson ProblemT1}(b), we present the radiation temperature (as obtained by different approximations), versus the exact solution. We can see that all approximation are bunched close to the IMC due to the fact that $T_H=1$ yields an optically thick problem.

\begin{figure}[htbp!]
\centering 
(a)
\includegraphics*[width=7.5cm]{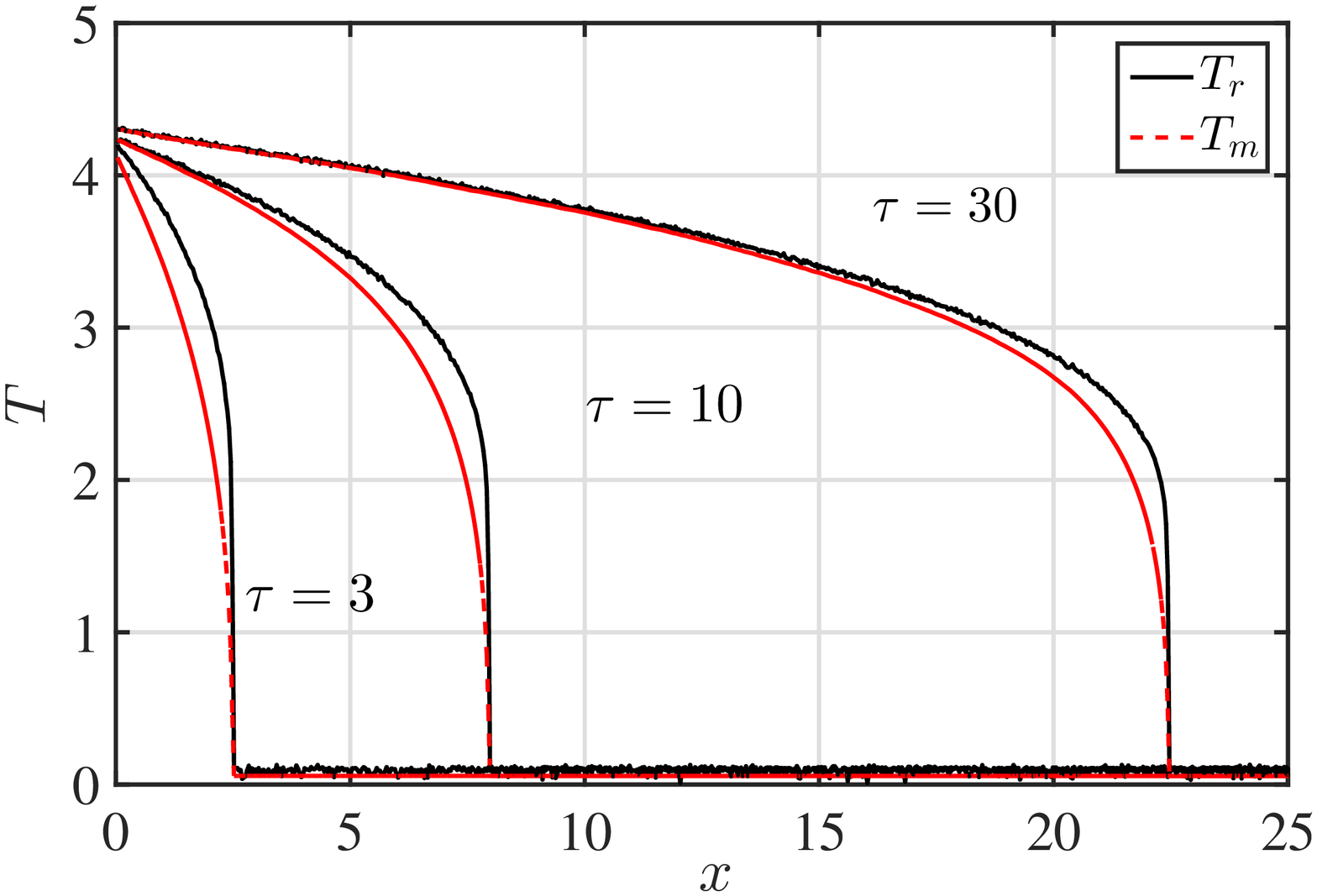}
(b)
\includegraphics*[width=7.5cm]{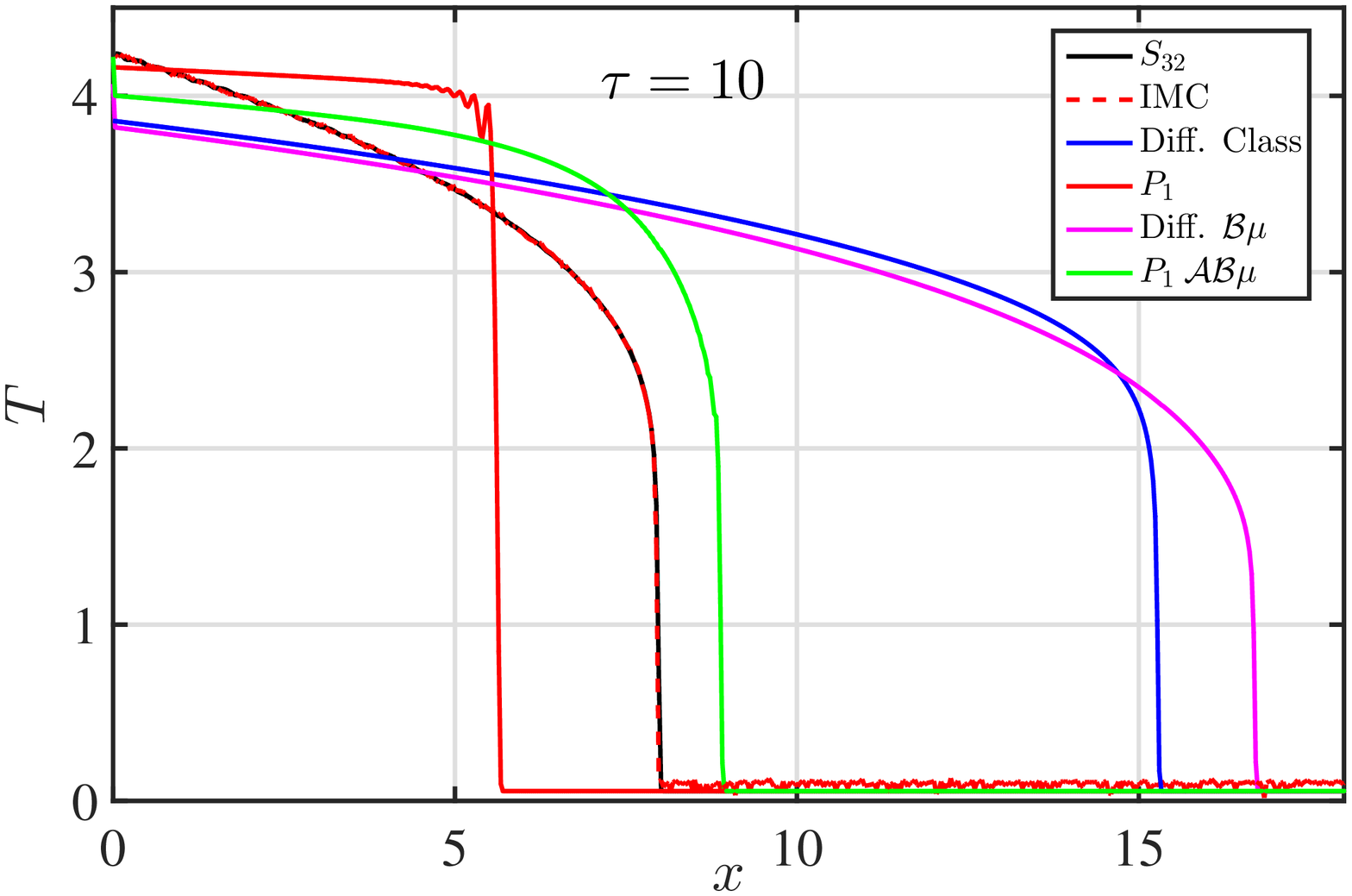}
\caption[NL_Olson ProblemT5]{(a) The radiation and material temperatures of the IMC in different times ($\tau=$3, 10, 30) for the Olson's-like problem using $T_H=5$. (b) The the radiation temperature ($T_r$), as a function of the optical depth in $\tau=10$. 
The red and blue solid curves are the classic diffusion and $P_1$. The magenta and the green curves are the $\B\mu$ diffusion and the $\A\B\mu$ $P_1$ approximations, respectively. The exact IMC is the black solid curves.}  
\label{fig:NL_Olson ProblemT5}
\end{figure}

Thus, we offer an Olson's-like optically thin benchmark, by increasing the incoming flux and set $T_H=5$ (and thus, $F_{in}=\frac{5acT_{H}}{4}$). Since the opacity of the problem depends as $T_m^{-3}$ with the material temperature (Eq.~\ref{NLOlsonOpacity}), the opacity decreases significantly. In Fig.~\ref{fig:NL_Olson ProblemT5} the results of the Olson-like nonlinear opacity benchmark using $T_H=5$ are shown. We can see in Fig.~\ref{fig:NL_Olson ProblemT5}(a) that the difference between the radiation and material temperatures in different times increases in comparison with $T_H=1$ case. Moreover, In Fig.~\ref{fig:NL_Olson ProblemT5}(b) we introduce the radiation temperature using several approximations and the exact (IMC) solution. We can see that the $P_1$ (red solid curve) is too slow, and both the classic and Zimmerman's $\mu\B$ diffusion approximations (blue and magenta solid curved) propagate too fast. 
The new $\mu\A\B$ approximation yields quite close results to the exact solutions, obtaining almost the correct heat front. 

\section{Energy density and flux discontinuity ($\alpha\beta\B$ and $\alpha\beta\A\B$ Approximations)}
\label{alpha_beta_sec}

In Sec.~\ref{s6} we have introduced the two-region Milne problem, indicating that both the asymptotic energy density and flux are discontinuous. 
In Sec.~\ref{s6b} we noted that Zimmerman offered a Marshak-like approximation for the jump conditions that have discontinuity in the energy density but have a continuous flux (and thus, conserves particles). Next, in Sec.~\ref{ours} we introduced the new $\mu\A\B$ approximation that uses Zimmerman's Marshak-like approximate jump conditions to derive a modified discontinuous asymptotic $P_1$ approximation. The question we now wish to pose is whether we can go further and employ the precise Milne jump conditions to derive an even more accurate approximation.

First, McCormick et. al. solved the exact two-region problem, finding the discontinuous jump conditions  
of the energy density ($\rho_{\nicefrac{2}{1}}=E^A_{\mathrm{as}}/E^B_{\mathrm{as}}$) and the flux ($j_{\nicefrac{2}{1}}=\vec{F}^A_{\mathrm{as}}/\vec{F}^B_{\mathrm{as}}$)~\cite{mccormick1,mccormick2}. both $\rho_{\nicefrac{2}{1}}$ and $j_{\nicefrac{2}{1}}$ are functions of the $\omega_{\mathrm{eff}}(\vec{r},t)$ of the two media, $\omega_{\mathrm{eff}}^A$ and $\omega_{\mathrm{eff}}^B$. McCormick at al. have also fully tabulated the numerical values of $\rho_{\nicefrac{2}{1}}$ and $j_{\nicefrac{2}{1}}$~\cite{mccormick3}. We note that the exact solution of the two-region problem was introduced in many other papers, for example in~\cite{ganapol_pomraning}. A minor approximation, based on variational analysis yields very close values of the discontinuities, by introducing the discontinuities in both energy density and radiation flux as~\cite{rul_pom_su,ganapol_pomraning}:
\begin{subequations}
\begin{equation}
\beta_A(\vec{r_S},t)E_A(\vec{r_S},t)=\beta_B(\vec{r_S},t)E_B(\vec{r_S},t)
\label{dis_fj1}
\end{equation}
\begin{equation}
\alpha_A(\vec{r_S},t)\vec{F}_A(\vec{r_S},t)=\alpha_B(\vec{r_S},t)\vec{F}_B(\vec{r_S},t)
\label{dis_fj2}
\end{equation}
\label{dis_fj}
\end{subequations}
The dependence of $\alpha(\vec{r_S},t)$ and $\beta(\vec{r_S},t)$ in space and time is again due to $\omega_{\mathrm{eff}}$ (see Appendix A). This form of applying the discontinuous condition is more convenient to apply in numerical codes, setting $\rho_{\nicefrac{2}{1}}=\beta(\omega_{\mathrm{eff}}^B)/\beta(\omega_{\mathrm{eff}}^A)$ and $j_{\nicefrac{2}{1}}=\alpha(\omega_{\mathrm{eff}}^B)/\alpha(\omega_{\mathrm{eff}}^A)$, and the difference from the exact solution is minor (for an accuracy check comparing to the exact McCormick solutions, see Appendix C).

Following the procedure described in Zimmerman's discontinuous diffusion (Sec.~\ref{s6b}), Eqs.~\ref{dis_fj} yields modified $P_1$ equations (see in~\cite{rul_pom_su,ganapol_pomraning} for the time-independent case): 
\begin{subequations}
\begin{equation}
 \frac{1}{c}\frac{\partial E(\vec{r},t)}{\partial t}+\frac{1}{c\alpha(\vec{r},t)}\nabla\cdot \left( 
 \alpha(\vec{r},t) \vec{F}(\vec{r},t) \right)
 =\sigma_{a}(T_m(\vec{r},t))\left(\int_{4\pi}\frac{B(\vec{r},t)}{c}-E(\vec{r},t)\right)+\frac{S(\vec{r},t)}{c}
\label{DisC1}
\end{equation}
\begin{equation}
\vec{F}(\vec{r},t)=-\frac{cD(\vec{r},t)}{\beta(\vec{r},t)}\vec{\nabla}\left(\beta(\vec{r},t)E(\vec{r},t)\right),
\label{ZFick2}
\end{equation}
\label{diff_ab}
\end{subequations}
Eq.~\ref{DisC1} replaces the conservation law (Eq.~\ref{Rad1}), and thus {\em does not} conserves particles (the conserved quantity is $\alpha(\vec{r},t) \vec{F}(\vec{r},t)$ instead), which makes it less favorable. Eq.~\ref{ZFick2} is identical to Eq.~\ref{ZFick}, replacing $\beta(\omega_{\mathrm{eff}})$ with $\mu(\omega_{\mathrm{eff}})$. Eqs.~\ref{diff_ab} yields a discontinuous asymptotic diffusion, that does not conserves particles. By recalling that $\B(\omega_{\mathrm{eff}})=1/D_0(\omega_{\mathrm{eff}})$, this diffusion approximation is called the $\alpha\beta\B$ approximation. 

Next, in a similar way to the derivation of the new $\mu\A\B$ approximation (see Sec.~\ref{ours}), we can derive a modified $\alpha\beta\A\B$ $P_1$ equation, Eq.~\ref{DisC1} and:
\begin{equation}  
\beta(\vec{r},t)\frac{\A(\vec{r},t)}{c}\frac{\partial F(\vec{r},t)}{\partial t}+c\vec{\nabla}\left({\beta(\vec{r},t)}E(\vec{r},t)\right)+
\beta(\vec{r,t})\B(\vec{r},t){\sigma_{t}((T_m(\vec{r},t))}F(\vec{r},t)=0
\label{DisC2beta}
\end {equation}
which is identical to Eq.~\ref{DisC2}, replacing $\beta(\omega_{\mathrm{eff}})$ with $\mu(\omega_{\mathrm{eff}})$. Eqs.~\ref{DisC1} and~\ref{DisC2beta} are thus the $\alpha\beta\A\B$ approximation.

\begin{figure}[htbp!]
\centering 
(a)
\includegraphics*[width=7.5cm]{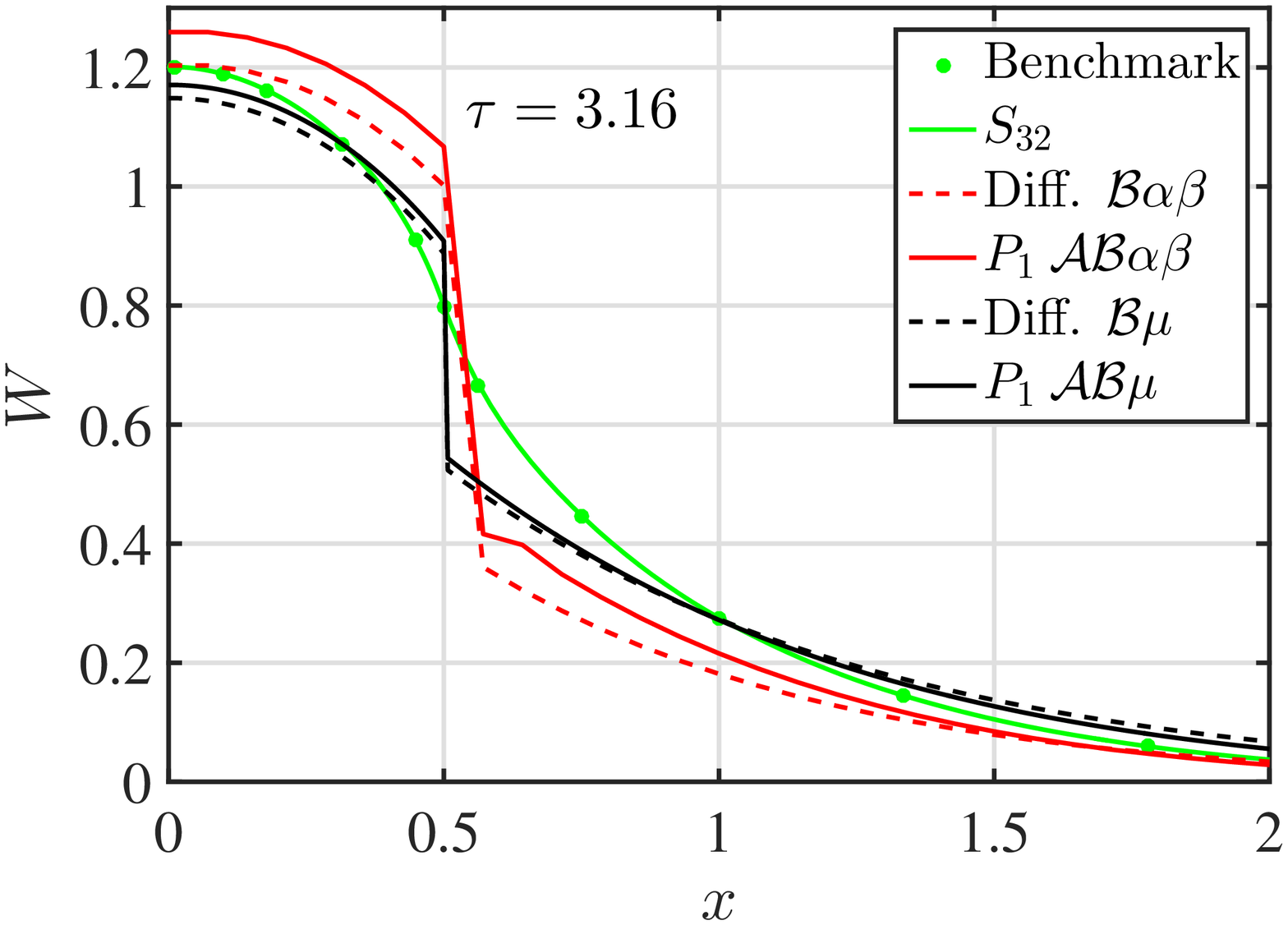}
(b)
\includegraphics*[width=7.5cm]{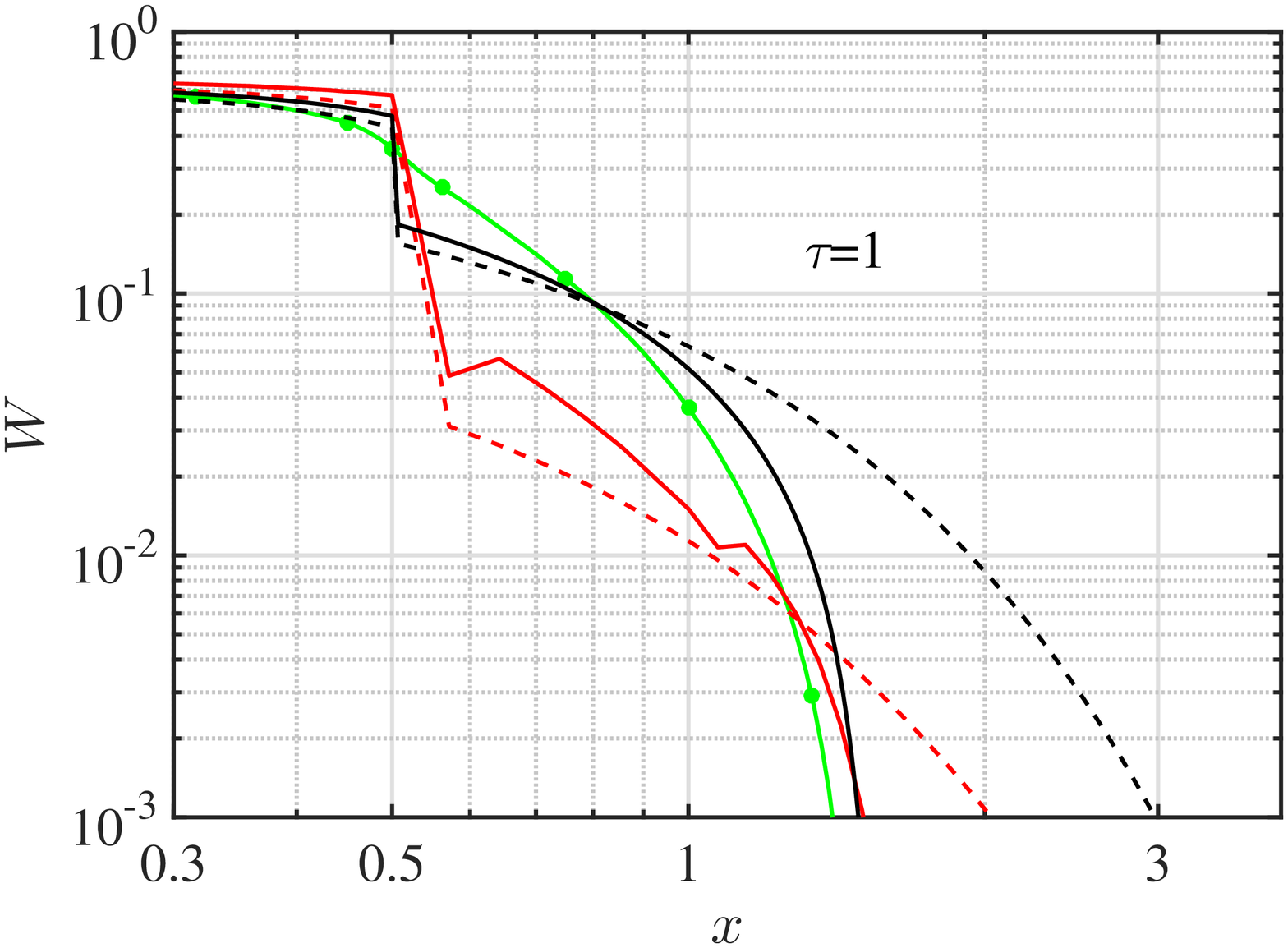}
\caption[SuOlson Problem AlpBet1]{The radiation energy density (W) in linear (a) and logarithmic (b) scales as function
of space in different times, for the case of $c_s=0$. The exact transport solution is in The circles are where the $S_{32}$
are represented by the green curves. The red dashed and solid curves are the $\alpha\beta\B$ discontinuous diffusion and the $\A\B\alpha\beta$ discontinuous $P_1$ approximations. 
The blue solid and dashed curves are the $\B\mu$ discontinuous diffusion and $\A\B\mu$ discontinuous $P_1$ approximations.}  
\label{fig:SuOlsonAlphBetCa1p0AB}
\end{figure}

\begin{figure}[htbp!]
\centering 
(a)
\includegraphics*[width=7.5cm]{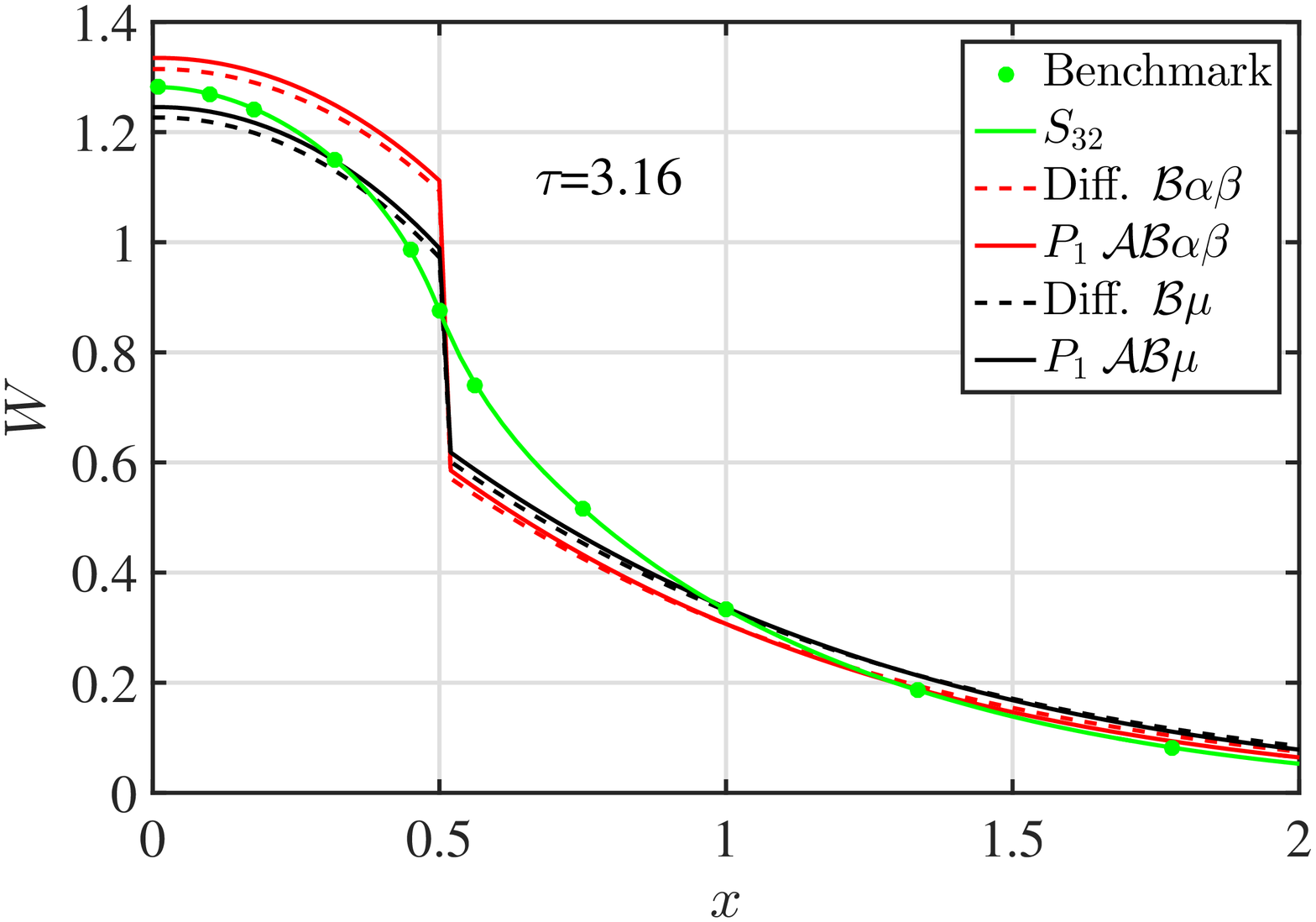}
(b)
\includegraphics*[width=7.5cm]{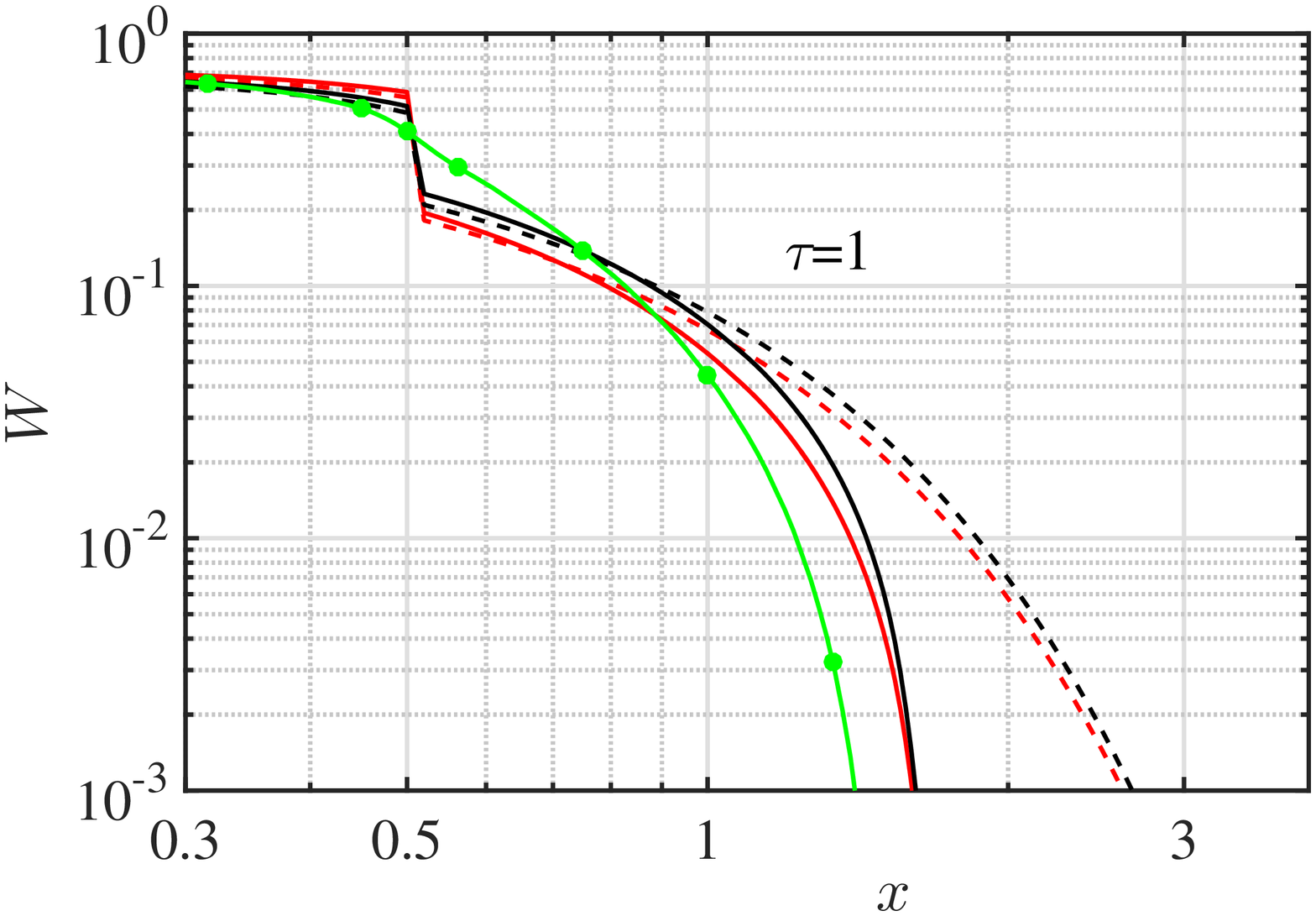}
\caption[SuOlson Problem AlpBet0p5]{The same as Fig.~\ref{fig:SuOlsonAlphBetCa1p0AB} for the scattering-included case, $c_s=0.5$.}  
\label{fig:SuOlsonAlphBetCa0p5AB}
\end{figure}

The results of the Su-Olson constant opacity benchmark using this $\alpha\beta\A\B$ approximation (in discontinuous $P_1$ notation) and the $\alpha\beta\B$ approximation (in discontinuous diffusion notation) are presented in Fig.~\ref{fig:SuOlsonAlphBetCa1p0AB} for $c_s=0$, and in Fig.~\ref{fig:SuOlsonAlphBetCa0p5AB} for $c_s=0.5$. 

First, it turns out that using $\alpha(\omega_{\mathrm{eff}})$ and $\beta(\omega_{\mathrm{eff}})$ instead of $\mu(\omega_{\mathrm{eff}})$, causes essential numerical difficulties, especially in the purely absorbing case (which is the most common physical case; scattering is usually negligible). The noisy results can be seen in the purely absorbing case in
Fig.~\ref{fig:SuOlsonAlphBetCa1p0AB}, and the numerical scheme is often unstable. This is due to the fact that 
$\alpha(\omega_{\mathrm{eff}})$ and $\beta(\omega_{\mathrm{eff}})$ both go to infinity when $\omega_{\mathrm{eff}}\to 0$~\cite{rul_pom_su,ganapol_pomraning}.
In the scattering-included case, $c_s=0.5$, the results are much smoother as can be seen in Fig.~\ref{fig:SuOlsonAlphBetCa0p5AB}, since the scattering prevents the $\omega_{\mathrm{eff}}\to 0$ limit. We note again, as is the case in any diffusion approximation, the $\alpha\beta\B$ approximation yields a heat front that is too fast.
 When the approximations are stable (such as the scattering-included case), the results have similar (or less) accuracy as the new $\mu\A\B$ approximation.

In conclusion, since the fact that in many cases this approximation is numerically unstable, and when the solution is available the accuracy is similar to (or even less than) the stable $\mu\A\B$ approximations, we {\em do not} recommend using $\alpha\beta\A\B$ or $\alpha\beta\B$ approximations (at least in radiative transfer problems).

\section{Discussion}
\label{discussion}
In this paper we have derived a new approximate method for solving the mono-energetic gray transport equation, the discontinuous asymptotic $P_1$ approximation (or the $\mu\A\B$ approximation). This method rests on two foundations: The asymptotic $P_{1}$ approximation~\cite{Heizler2010}, that reproduces the asymptotic steady-state behavior and prevents the infinite particle velocities (unlike the diffusion
approximations), and the discontinuity jump conditions of Zimmerman's discontinuous diffusion~\cite{zimmerman1979}, forcing a discontinuity in the energy density and continuous flux (and thus, conserves particles).

We show that this approximation yields better results than do other common methods in two important benchmark problems, the Su-Olson constant opacity benchmark (both with or without scattering)~\cite{SuOlson1996} and Olson's nonlinear opacity (temperature-dependent) problem~\cite{Olson1999}. The new approximation yields even better results than the gradient-dependent approximations, such as various Flux-Limiter approximations or the variable Eddington factor approximations. We consider this method to be better grounded in physics than others, in that it relies on precise asymptotic solutions, which are indeed discontinuous. That may explain the quality of its results.

We have  also tested the possibility for using a method that includes discontinuities in both energy density and radiation flux (the $\alpha\beta\A\B$ approximation), based on the exact two-region Milne problem. We have found that these methods often suffer from numerical instabilities, while when stable the accuracy is similar to the $\mu\A\B$ approximation. Due to these observations, and the fact that this approximation does not conserves particles, we conclude that the $\mu\A\B$ approximation is preferable.

In future work, we plan to test the new approximation against actual supersonic Marshak-wave experiments~\cite{Back2000,Moore2015}, comparing it to exact approaches such as $S_N$ or IMC. In addition, it would be interesting to test the new approximation in 2D/3D. The new method depends explicitly {\em only} on $\omega_\mathrm{eff}$, when  $\omega_\mathrm{eff}$
is defined on the middle of the numerical cell, just like $E$. In gradient-dependent approximations such as the VEF or FL, the approximation depends 
on $\vec{F}/E$, where $\vec{F}$ is defined on cell edges, which makes it much more complicated to solve in multi-dimensional scheme

This numerical advantage of the new scheme will become very important if it can be extended to higher dimensions.

\appendix
\section{Numerical Values For $\A(\omega_\mathrm{eff})$, $\B(\omega_\mathrm{eff})$ and $\mu(\omega_\mathrm{eff})$}
\label{numrical_expressions}

Here we introduce full numerical expressions that were used 
for the $\omega$-dependent functions (For simplicity, we set here $\omega_\mathrm{eff}\equiv\omega$):
We recall that $\B(\omega)$ in Eq.~\ref{Rad2H} is equal to $=1/D_0(\omega)$ from
Eq.~\ref{Asymptotic diffusion}. $\A(\omega)$ and $\B(\omega)$ were taken as was explained in~\cite{Heizler2010,Heizler2012,Ravetto_Heizler2012}:
\begin{equation}
\A(\omega)=\begin{cases}
     0.96835-0.437\omega, &\text{if} \, 0.55 \leq \omega \leq 0.65
		\\\frac{0.247(0.433+0.421\omega- 2.681\omega^2- 1.82\omega^3+4.9\omega^4- 1.06\omega^5+2.56\omega^6)}{ 
		( 0.33+0.159\omega- 0.567\omega^2-\omega^3)^2} 
		, &   \text{otherwise}
   \end{cases}
\label{Aparam}
\end{equation}

\begin{equation}
\B(\omega)=\begin{cases}
     \frac{1}{0.80054- 0.523\omega}, & 
     \text{if} \, 0.59 \leq \omega \leq 0.61
    \\ \frac{0.1326495+\omega[0.03424169+\omega(0.1774006-\omega)] }{0.3267567+\omega[0.1587312-\omega(0.5665676+\omega)] }\cdot\frac{ 1+\omega }{0.40528473}, &   \text{otherwise}
   \end{cases}
\label{Bparam}
\end{equation}

Calculating the third $\omega$-dependent function, $\mu(\omega)$ as was defined in Eq.~\ref{muAsymptotic} is through the definition of $\kappa(\omega)$, the solution of the transcendental Eq.~\ref{kappa}. A numerical evaluation of $\kappa(\omega)$ can be~\cite{Case1953}:
 
\begin{equation}
\kappa^2(\omega)=\begin{cases}
    1, &   \omega<0.01
    \\ 1-4e^{-\frac{2}{\omega}}\left(1+\frac{4-2\omega}{\omega}e^{-\frac{2}{\omega}}+\frac{24+20\omega+3\omega^2}{\omega^2}e^{-\frac{4}{\omega}}\right) , & 0.01<\omega \leq 0.45
     \\ (1-\omega)\B(\omega), & 0.45<\omega<1 \\
      (\omega-1)\B(\omega), & \omega\ge 1
   \end{cases}
\label{kappa1}
\end{equation}

Subsequently, $\mu(\omega)$ itself is calculated~\cite{zimmerman1979}:
\begin{equation}
\mu(\omega)=\begin{cases}
    
    1, &   \omega<0.01
     \\  \begin{cases}  -\frac{\omega}{2\kappa^2(\omega)}\log{(1-\kappa^2(\omega))}, & \kappa^2(\omega)>0 \\ 1,& \mathrm{otherwise} \end{cases}

     , & 0.01\le\omega < 0.999
     \\ \ln\left(\frac{8.3548+1.5708+\omega}{2.1228+2.4674\omega}\right), & 0.999 \leq \omega \leq 1.001
     \\ \frac{\omega}{2\kappa^2(\omega)}\log{(1+\kappa^2(\omega))}, & \omega > 1.001
     
   \end{cases}
\label{mu}
\end{equation}

\begin{figure}[htbp!]
\centering 
\includegraphics*[width=7.5cm]{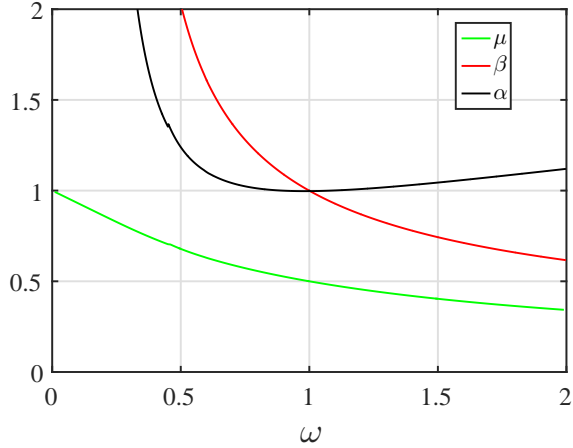}
\caption{The functions $\mu(\omega)$, $\beta(\omega)$ and $\alpha(\omega)$. $\beta(\omega)$ and $\alpha(\omega)$ goes to infinity when $\omega\to 0$.}
\label{fig:betaonly}
\end{figure}

$\alpha(\omega)$ and $\beta(\omega)$ from Eqs.~\ref{dis_fj}, were calculated in a manner similar to that suggested in~\cite{rul_pom_su} (Eqs.~95-96) or in~\cite{ganapol_pomraning} (Eqs.~77-78). In Fig.~\ref{fig:betaonly} we introduce the curves of $\mu(\omega)$, $\beta(\omega)$ and $\alpha(\omega)$. We can see that both $\beta(\omega)$ and $\alpha(\omega)$ go to infinity when $\omega\to 0$, casing numerical instabilities on the $\alpha\beta$-included approximations. Similar figures for $\A(\omega)$ and $\B(\omega)$ may be found in~\cite{Heizler2010,Heizler2012,Ravetto_Heizler2012}.

\section{The Discontinuous Asymptotic $P_1$ Approximation for Neutronics}
\label{neutronics}

In neutronics, the mono-energetic Boltzmann equation is (equivalent to Eq.~\ref{Boltz} in this work)~\cite{Heizler2012}:
\begin{align}
\label{BoltzmanMonoN}
& \frac{1}{v}\frac{\partial \psi(\hat{\Omega},\vec{r},t)}{\partial t}+
\hat{\Omega}\cdot \vec{\nabla} \psi(\hat{\Omega},\vec{r},t)+\Sigma_t(\vec{r})
\psi(\hat{\Omega},\vec{r},t)= \\
& \int_{4\pi}{d\hat{\Omega'}\Sigma_s(\hat{\Omega} \cdot \hat{\Omega'},\vec{r})
\psi(\hat{\Omega},\vec{r},t)}+\nu(\vec{r})\Sigma_f(\vec{r})\int_{4\pi}{d\hat{\Omega'}
\psi(\hat{\Omega'},\vec{r},t)}+S(\hat{\Omega},\vec{r},t)\nonumber
\end{align}
when $\psi(\hat{\Omega},\vec{r},t)$ is the angular flux. $\Sigma_t(\vec{r})=\Sigma_a(\vec{r})+\Sigma_s(\vec{r})+\Sigma_f(\vec{r})$ is the total cross-section when $\Sigma_a(\vec{r})$ is the absorbing cross-section, $\Sigma_s(\vec{r})\equiv \int_{-1}^1{d\mu_0\Sigma_s(\mu_0),\vec{r})}$ is the scattering cross-section ($\mu_0\equiv \hat{\Omega} \cdot \hat{\Omega'}$) and $\Sigma_f(\vec{r})$ is the fission cross-section. $S(\hat{\Omega},\vec{r},t)$ is an external source term, $\nu(\vec{r})$ is the mean number of neutrons that are emitted per fission and $v$ is the neutron velocity. Here we use the scalar flux $\phi(t,\vec{r})$ and the total current $\vec{J}(t,\vec{r})$ as the first two moments of $\psi(\hat{\Omega},\vec{r},t)$ (equivalent to $E(\vec{r},t)$ and $\vec{F}(\vec{r},t)$ in this work), while $c(t,\vec{r})$ (which is called {\em the albedo}), the mean number of particles emitted from a collision, is replacing $\omega_\mathrm{eff}(t,\vec{r})$, and is defined as:
\begin{equation}
c(\vec{r},t)=\frac{\Sigma_s(\vec{r})+\nu(\vec{r})\Sigma_f(\vec{r})+S(t,\vec{r})/\left(v\Sigma_t(\vec{r})\phi(t,\vec{r})\right)}{\Sigma_t(\vec{r})}
\label{albedo}
\end{equation}
The first $P_1$ equation, the conservation law for neutronics is (equivalent to Eq.~\ref{Rad1}): 
\begin{equation}
\frac{1}{v}\frac{\partial \phi(t,\vec{r})}{\partial t}+\vec{\nabla} \cdot
\vec{J}(t,\vec{r})+\Sigma_a(\vec{r})\phi(t,\vec{r})=(\nu(\vec{r})-1)\Sigma_f(\vec{r})\phi(t,\vec{r})+S(t,\vec{r}),
\label{conserv_p}
\end{equation}
and the equivalent to the second discontinuous asymptotic $P_1$ equation, Eq.~\ref{DisC2}, is:
\begin{equation}  
\mu(\vec{r},t)\frac{\A(\vec{r},t)}{v}\frac{\partial \vec{J}(\vec{r},t)}{\partial t}+\vec{\nabla}\left({\mu(\vec{r},t)}\phi(\vec{r},t)\right)+
\mu(\vec{r,t})\B(\vec{r},t)\Sigma_{t}((\vec{r})\vec{J}(\vec{r},t)=0.
\label{DisC2nu}
\end {equation}

\section{The Accuracy of $\alpha(\omega_\mathrm{eff})$ and $\beta(\omega_\mathrm{eff})$ Discontinuity Jump Conditions}
\label{mccormick_vs_ab}

In this appendix we introduce the accuracy of using the approximate variational analysis of the discontinuity jump condition that was introduced in~\cite{rul_pom_su,ganapol_pomraning}, comparing to the exact numerical two-region Milne problem solutions~\cite{mccormick3}.

\begin{figure}[htbp!]
\centering 
(a)
\includegraphics*[width=7.5cm]{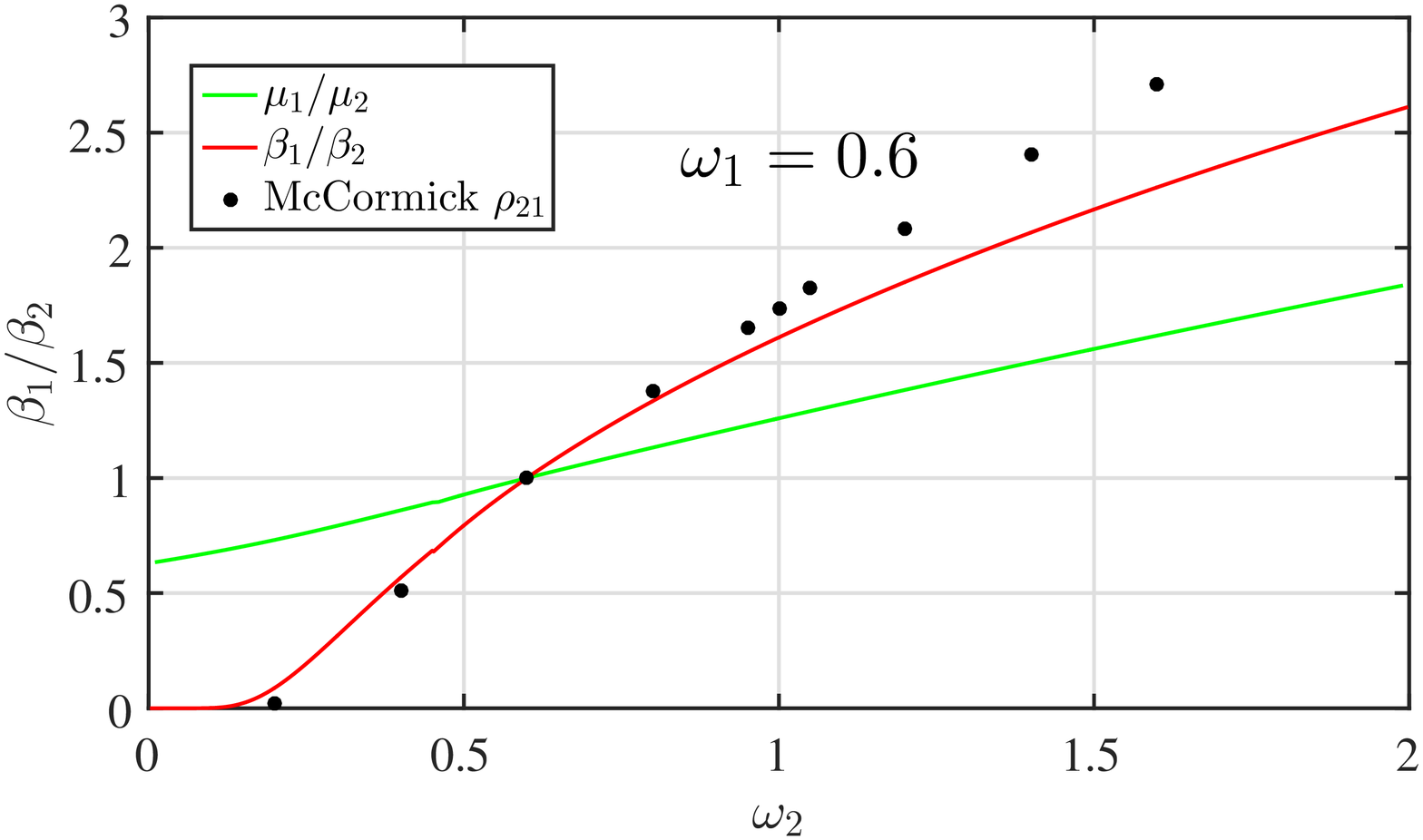}
(b)
\includegraphics*[width=7.5cm]{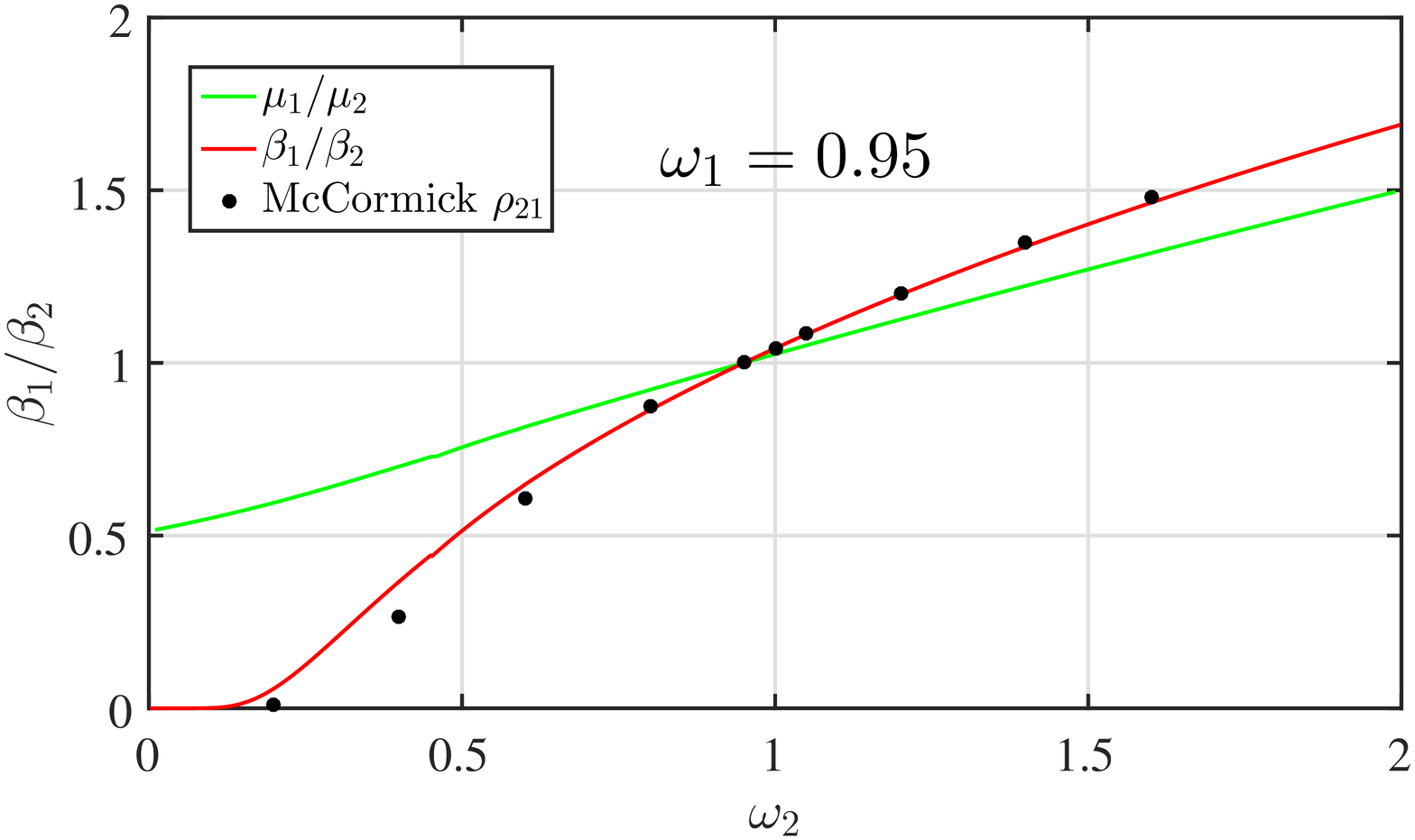}
\caption[beta]{The ratio $\beta_1/\beta_2$, between two
region, as a function of $\omega_2$ using the exact McCormick solution~\cite{mccormick3} (black circles) and the approximate values of the variational analysis~\cite{rul_pom_su,ganapol_pomraning} (red curves) for
$\omega_1=0.6$ (a), and for $\omega_1=0.95$ (b).}
\label{fig:beta}
\end{figure}

\begin{figure}[htbp!]
\centering 
(a)
\includegraphics*[width=7.5cm]{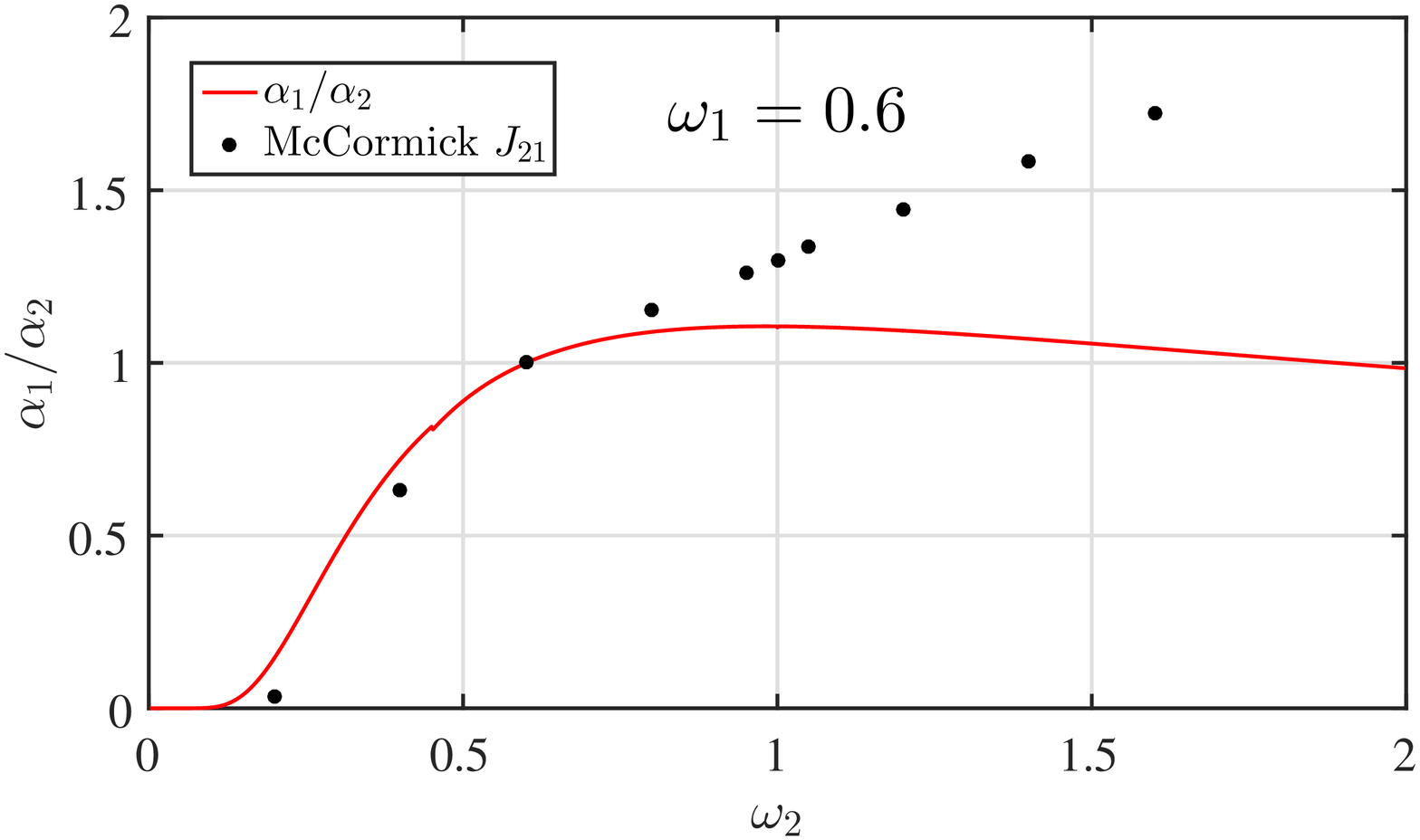}
(b)
\includegraphics*[width=7.5cm]{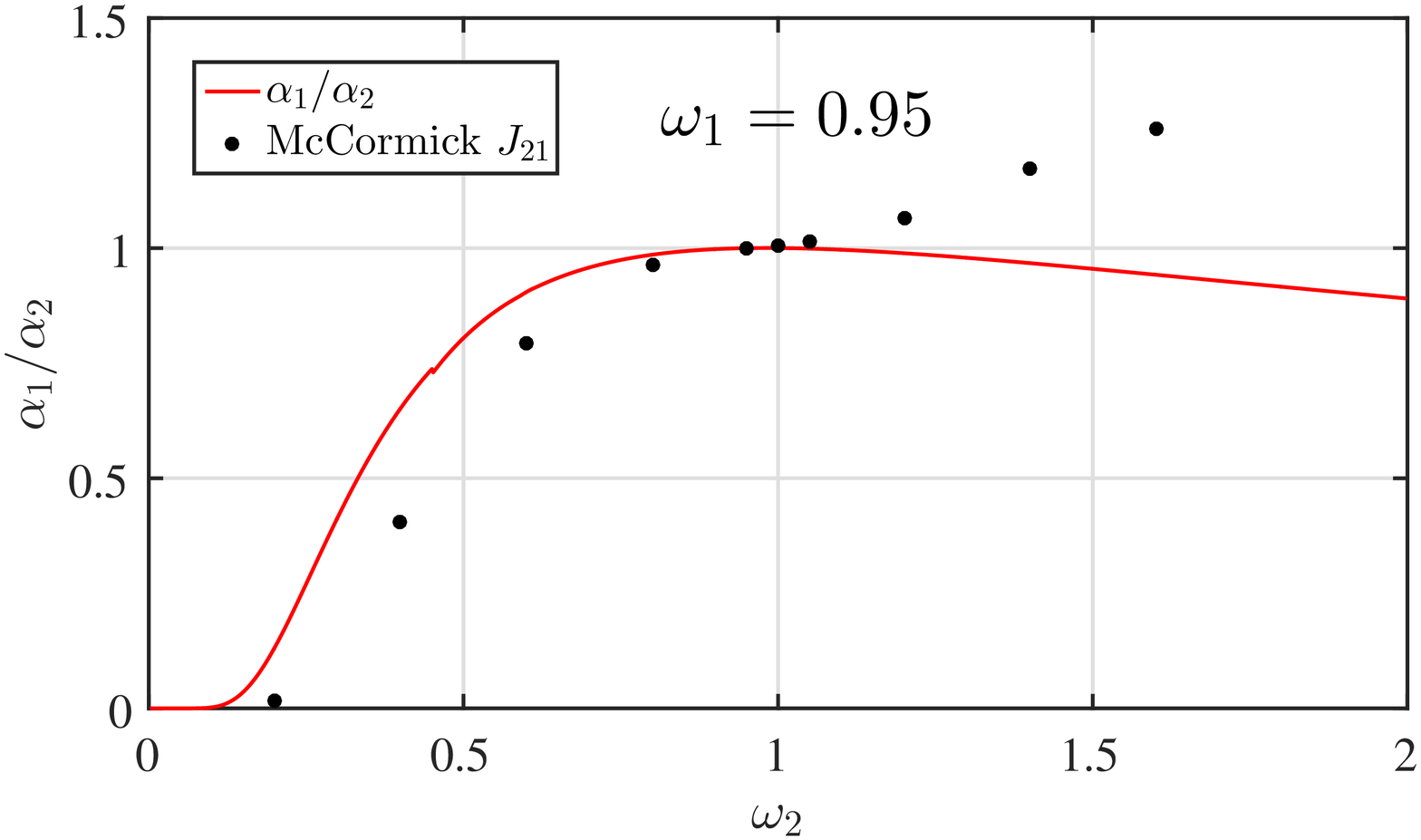}
\caption[alpha]{The ratio $\alpha_1/\alpha_2$, between two
region, as a function of $\omega_2$ using the exact McCormick solution~\cite{mccormick3} (black circles) and the approximate values of the variational analysis~\cite{rul_pom_su,ganapol_pomraning} (red curves) for
$\omega_1=0.6$ (a), and for $\omega_1=0.95$ (b).}
\label{fig:alpha}
\end{figure}

The exact energy density discontinuity $\rho_{\nicefrac{2}{1}}$ is compared with the approximated $\beta_1/\beta_2$ (Fig.~\ref{fig:beta}, along with Zimmerman's $\mu_1/\mu_2$) and the exact flux discontinuity $j_{\nicefrac{2}{1}}$ is compared to $\alpha_1/\alpha_2$ (Fig.~\ref{fig:alpha}) as a function of $\omega_2$ for two numerical values of $\omega_1$, 0.6 (a) and 0.95 (b).

We can see that the ratio between the zero moments ($E(\vec{r},t)$) $\beta_1/\beta_2$ fits quite well along all the range to exact McCormick calculations. One should remember that $\mu_1/\mu_2$ (from Zimmerman's approximation) should not suppose to be similar to $\beta_1/\beta_2$, since in the $\mu\A\B$ approximation we keep {\em only} the flux continuous (forcing $\alpha_1/\alpha_2=1$). The ratio between the first moments ($\vec{F}(\vec{r},t)$) $\alpha_1/\alpha_2$ shows that when $\omega_2<1$, $\alpha_1/\alpha_2$ is also similar to the exact McCormick calculations, while for $\omega_2>1$, the accuracy decreases. However, the total accuracy of the approximate variational analysis to the exact solutions, is quite good. 

In any case, the decrease of the McCormick exact $\rho_{\nicefrac{2}{1}}$ and $j_{\nicefrac{2}{1}}$, or the approximate variational analysis
values, to zero when $\omega_2\to 0$, makes it often numerically unstable, making the $\mu\A\B$ a preferable choice.

\begin{acknowledgments}
We acknowledge the support of the PAZY Foundation under Grant No.~61139927.
The authors thank Roee Kirschenzweig for using an IMC code for radiative problems, Stanislav Burov and the anonymous referees for their valuable comments.
\end{acknowledgments}

\end{document}